\documentclass[preprint,showpacs,preprintnumbers,amsmath,amssymb,nofootinbib,superscriptaddress]{revtex4}

\usepackage{graphicx}
\usepackage{dcolumn}
\usepackage{bm}
\usepackage{units}
\newcommand{\Cambridge}{Cavendish Laboratory, University of Cambridge, Madingley Road, Cambridge CB3 0HE, United Kingdom}
\newcommand{\FNAL}{Fermi National Accelerator Laboratory, Batavia, Illinois 60510, USA}
\newcommand{\RAL}{Rutherford Appleton Laboratory, Science and Technology Facilities Council, OX11 0QX, United Kingdom}
\newcommand{\UCL}{Department of Physics and Astronomy, University College London, Gower Street, London WC1E 6BT, United Kingdom}
\newcommand{\Caltech}{Lauritsen Laboratory, California Institute of Technology, Pasadena, California 91125, USA}
\newcommand{\ANL}{Argonne National Laboratory, Argonne, Illinois 60439, USA}
\newcommand{\Athens}{Department of Physics, University of Athens, GR-15771 Athens, Greece}

\newcommand{\Benedictine}{Physics Department, Benedictine University, Lisle, Illinois 60532, USA}
\newcommand{\BNL}{Brookhaven National Laboratory, Upton, New York 11973, USA}

\newcommand{\Delhi}{Department of Physics and Astrophysics, University of Delhi, Delhi 110007, India}

\newcommand{\Harvard}{Department of Physics, Harvard University, Cambridge, Massachusetts 02138, USA}
\newcommand{\HolyCross}{Holy Cross College, Notre Dame, Indiana 46556, USA}
\newcommand{\IIT}{Physics Division, Illinois Institute of Technology, Chicago, Illinois 60616, USA}
\newcommand{\Indiana}{Indiana University, Bloomington, Indiana 47405, USA}

\newcommand{\Lebedev}{Nuclear Physics Department, Lebedev Physical Institute, Leninsky Prospect 53, 119991 Moscow, Russia}
\newcommand{\LLL}{Lawrence Livermore National Laboratory, Livermore, California 94550, USA}

\newcommand{\Minnesota}{University of Minnesota, Minneapolis, Minnesota 55455, USA}

\newcommand{\Duluth}{Department of Physics, University of Minnesota -- Duluth, Duluth, Minnesota 55812, USA}
\newcommand{\Otterbein}{Otterbein College, Westerville, Ohio 43081, USA}
\newcommand{\Oxford}{Subdepartment of Particle Physics, University of Oxford, Oxford OX1 3RH, United Kingdom}
\newcommand{\Penn}{Physics and Astronomy, University of Pennsylvania, Philadelphia, Pennsylvania 19104, USA}
\newcommand{\Pittsburgh}{Department of Physics and Astronomy, University of Pittsburgh, Pittsburgh, Pennsylvania 15260, USA}
\newcommand{\IHEP}{Institute for High Energy Physics, Protvino, Moscow Region RU-140284, Russia}

\newcommand{\Carolina}{Department of Physics and Astronomy, University of South Carolina, Columbia, South Carolina 29208, USA}

\newcommand{\Stanford}{Department of Physics, Stanford University, Stanford, California 94305, USA}
\newcommand{\StJohnFisher}{Physics Department, St. John Fisher College, Rochester, New York 14618 USA}
\newcommand{\Sussex}{Department of Physics and Astronomy, University of Sussex, Falmer, Brighton BN1 9QH, United Kingdom}
\newcommand{\TexasAM}{Physics Department, Texas A\&M University, College Station, Texas 77843, USA}
\newcommand{\Texas}{Department of Physics, University of Texas at Austin, 1 University Station C1600, Austin, Texas 78712, USA}

\newcommand{\Tufts}{Physics Department, Tufts University, Medford, Massachusetts 02155, USA}
\newcommand{\UNICAMP}{Universidade Estadual de Campinas, IF-UNICAMP, CP 6165, 13083-970, Campinas, SP, Brazil}
\newcommand{\USP}{Instituto de F\'{i}sica, Universidade de S\~{a}o Paulo,  CP 66318, 05315-970, S\~{a}o Paulo, SP, Brazil}
\newcommand{\Warsaw}{Department of Physics, University of Warsaw, Ho\.{z}a 69, PL-00-681 Warsaw, Poland}
\newcommand{\Washington}{Physics Department, Western Washington University, Bellingham, Washington 98225, USA}
\newcommand{\WandM}{Department of Physics, College of William \& Mary, Williamsburg, Virginia 23187, USA}
\newcommand{\Wisconsin}{Physics Department, University of Wisconsin, Madison, Wisconsin 53706, USA}
\newcommand{\deceased}{Deceased.}

\newcommand{\flux}{\ensuremath{\Phi^{\nu(\bar{\nu})}(E)}}
\newcommand{\ncc}{\ensuremath{N_\mathrm{CC}^{\nu(\bar{\nu})}(E)}}
\newcommand{\gcc}{\ensuremath{\Gamma_\mathrm{CC}^{\nu(\bar{\nu})}(E)}}
\newcommand{\acc}{\ensuremath{A_\mathrm{CC}^{\nu(\bar{\nu})}(E)}}
\newcommand{\bcc}{\ensuremath{B_\mathrm{CC}^{\nu(\bar{\nu})}(E)}}
\newcommand{\fsam}{\ensuremath{F^{\nu(\bar{\nu})}(E)}}
\newcommand{\aflux}{\ensuremath{A_{\Phi}^{\nu(\bar{\nu})}(E)}}
\newcommand{\bflux}{\ensuremath{B_{\Phi}^{\nu(\bar{\nu})}(E)}}
\newcommand{\nucor}{\ensuremath{S^{\nu(\bar{\nu})}(\nu_0,E)}}
\newcommand{\hecor}{\ensuremath{H^{\nu}}}
\newcommand{\xsec}{\ensuremath{\sigma_{CC}^{\nu(\bar{\nu})}(E)}}
\newcommand{\nub}{\ensuremath{\overline{\nu}}}
\begin{document}

\preprint{FERMILAB-PUB-09-468-E}

\title{Neutrino and Antineutrino Inclusive Charged-current Cross
  Section Measurements with the MINOS Near Detector}

\begin{abstract}
The energy dependence of the neutrino-iron and antineutrino-iron
inclusive charged-current cross sections and their ratio
have been measured using a high-statistics sample with the MINOS Near Detector
exposed to the NuMI beam from the Main Injector at Fermilab. 
Neutrino and antineutrino fluxes were determined using a low hadronic 
energy subsample of charged-current events. We report measurements of
$\nu$-Fe (\nub-Fe) cross section in the energy range 3-\unit[50]{GeV} (5-\unit[50]{GeV}) 
with precision of 2-8\% (3-9\%) and their ratio which is measured
with precision 2-8\%.
The data set spans the region from low energy,
where accurate measurements are sparse, up to the
high-energy scaling region where the cross section is well understood.

\end{abstract}

\affiliation{\ANL}
\affiliation{\Athens}
\affiliation{\Benedictine}
\affiliation{\BNL}
\affiliation{\Caltech}
\affiliation{\Cambridge}
\affiliation{\UNICAMP}
\affiliation{\FNAL}
\affiliation{\Harvard}
\affiliation{\HolyCross}
\affiliation{\IIT}
\affiliation{\Indiana}
\affiliation{\UCL}
\affiliation{\Minnesota}
\affiliation{\Duluth}
\affiliation{\Otterbein}
\affiliation{\Oxford}
\affiliation{\Pittsburgh}
\affiliation{\RAL}
\affiliation{\USP}
\affiliation{\Carolina}
\affiliation{\Stanford}
\affiliation{\Sussex}
\affiliation{\TexasAM}
\affiliation{\Texas}
\affiliation{\Tufts}
\affiliation{\Warsaw}
\affiliation{\WandM}

\author{P.~Adamson}
\affiliation{\FNAL}

\author{C.~Andreopoulos}
\affiliation{\RAL}

\author{K.~E.~Arms}
\affiliation{\Minnesota}

\author{R.~Armstrong}
\affiliation{\Indiana}

\author{D.~J.~Auty}
\affiliation{\Sussex}


\author{D.~S.~Ayres}
\affiliation{\ANL}

\author{C.~Backhouse}
\affiliation{\Oxford}



\author{P.~D.~Barnes~Jr.}
\affiliation{\LLL}

\author{G.~Barr}
\affiliation{\Oxford}

\author{W.~L.~Barrett}
\affiliation{\Washington}





% ...

\collaboration{The MINOS Collaboration}
\noaffiliation
\date{\today}

\maketitle

\section{Introduction}

Neutrino-nucleon and antineutrino-nucleon charged-current ($\nu_{\mu}N$ CC and  $\bar{\nu}_{\mu}$N CC) inclusive cross
sections above \unit[30]{GeV} have been determined by 
several experiments~\citep{ccfr_90,ccfrr,cdhsw}
with a combined precision of 2\%~\citep{rpp}.
The measured cross sections at these energies have a linear dependence on energy, which
agrees well with the prediction of the Quark Parton Model (QPM)~\citep{qpm}.

At lower energies, the cross section is both less well measured and 
difficult to model due to overlapping contributions
from quasi-elastic processes ($\nu_\mu + n \rightarrow \mu^- + p $), resonance
excitation followed by subsequent decay,
and the onset of deeply-inelastic scattering (DIS).
This energy range is of particular interest to ongoing and future neutrino 
oscillation searches in MINOS, NO$\nu$A~\citep{Ayres:2002nm}, and T2K~\citep{Itow:2001ee}.
Most cross section measurements in the $E_{\nu}<$\unit[30]{GeV} range
~\citep{bnl,crs,ggm_ps,ggm_sps,ihep_itep,ihep_jinr,skat}
have uncertainties of the order of 10\%.
Recently, NOMAD~\citep{nomad} has measured the cross section 
down to \unit[2.5]{GeV} with a precision of better than 4\%.
However, this result relies on a particle production model tuned to
 data~\citep{Astier:2003rj} to predict the neutrino flux.
In this paper we present a measurement of the $\nu_{\mu}N$ CC cross section
with a precision from 2-8\%, covering the 
3-\unit[50]{GeV} energy range using the MINOS Near Detector. 
Our analysis uses a low hadronic energy subsample to determine
the flux shape~\citep{Mishra:1990ax,seligman}. 

Antineutrino-nucleon charged-current cross
sections in the $E_{\nu}<$\unit[30]{GeV} range
suffer from the same complications listed above and tend to be even less well measured.
Several experiments reported results~\citep{ggm_sps,bebc,ggm-ps-anu,ihep_jinr,ihep_itep}; however data
coverage in energy was sparse and these measurements 
typically have larger than 10\% uncertainty.
Our measurement has higher precision, with uncertainties 
which range from 3-9\%.

The $\bar{\nu}_{\mu}$N CC to $\nu_{\mu}$N CC cross section ratio,
$r=\sigma^{\overline{\nu}}/\sigma^{\nu}$,
has been measured with a combined precision of better than 1\% at high energies~\citep{seligman,nutev}
but only one dedicated
measurement~\citep{eichten} has been performed in the $E_{\nu}<$\unit[30]{GeV}  range.
Gargamelle~\citep{eichten} reports measurements of $r$ from 
1-\unit[10]{GeV} with precision of about 20\%.
Our result substantially adds both coverage and precision to the
determination of $r$.
The ratio is more precisely determined than either 
cross section measured separately
due to a partial cancellation of most systematic effects and a 
cancellation of the normalization uncertainty.

The results in this paper can be used to tune and improve 
neutrino interaction generator models~\citep{neugen,genie}.
For example, neutrino scattering data are required for the modeling of the axial vector 
contribution to the cross section~\cite{fleming}.
Also, the cross section ratio $r$ is particularly sensitive to 
the modeling of $xF_3$, the parity violating structure function,
which enters into the numerator and denominator
with opposite sign, and to the antiquark content of the nucleon, which 
contributes differently to neutrino and antineutrino scattering.
In addition, at \unit[5]{GeV}, about 70\% of our event sample has negative 
4-momentum transfer squared, $Q^2$,
of less than \unit[1.5]{GeV$^2$}. This large, low-$Q^2$ sample provides
model sensitivity to the low-$Q^2$ QCD contributions 
(higher order QCD, higher-twist, and target mass corrections) 
which are difficult to calculate.

\subsection*{Overview of the Analysis}

The $\nu_{\mu}$ CC and $\bar{\nu}_{\mu}$ CC total cross sections as a function 
of incoming neutrino energy $E$ are determined from 
the inclusive charged-current interaction rate and the incident neutrino flux.
A sample of CC events (``cross section sample'') is selected and
a subsample of these events with low hadronic energy (``flux sample'') is defined. 
A Monte Carlo simulation which includes detailed detector geometry and
response is used to correct the flux and cross section samples for
detector acceptance and smearing effects.

Neutrino and antineutrino differential cross sections,  $\frac{d \sigma^{\nu,\nub}}{d\nu}$,
approach the same constant
value, independent of energy, in the limit of low-$\nu$, where $\nu$
is the energy transfered to the hadronic system.
A method which exploits this feature is used to
determine the energy dependence of the flux 
from the flux sample, which is then normalized using the world 
average cross section value measured above
\unit[30]{GeV}. To accomplish this we make use of the full range of our data
sample, which overlaps with the high energy measurements in the 30-\unit[50]{GeV}
 region. This ``low-$\nu$'' method has been used previously at high 
energies~\citep{Mishra:1990ax,seligman} and here it is adapted to the $E<$\unit[30]{GeV} range.

The neutrino beam, detector and the Monte Carlo simulation of the experiment are described in Sec.~\ref{sec:Detector-and-beamline}.
Sec.~\ref{sec:analysis} describes the event sample selection and the methods for extracting the flux and the cross section.
A discussion of systematic uncertainties and results are given in Sec.~\ref{sec:Systematics} and \ref{sec:Results}, respectively.

\section{beamline and detector\label{sec:Detector-and-beamline}}

MINOS is a two-detector, long baseline neutrino oscillation experiment
using the NuMI (Neutrinos at Main Injector) neutrino beam
at Fermilab. The oscillation parameters are measured~\citep{prl_3e20,prd_1e20}
by comparing the $\nu_\mu$ energy spectra at the Near Detector
located at Fermilab and the Far Detector located \unit[734]{km} away
in the Soudan Mine in northern Minnesota. In this section we describe
the neutrino beam, the Near Detector, and the Monte Carlo simulation.
More detailed descriptions of the beamline and the
MINOS detectors are given elsewhere~\citep{nim_paper}.

\subsection{Neutrino Beam}

The NuMI neutrino beam is produced from \unit[120]{GeV} 
protons extracted in a \unit[10]{$\mu$s} spill from the  Main Injector which 
impinge on a graphite target, with a typical intensity for the data presented 
here of $2.2\times10^{13}$ protons on target (PoT) per spill. 
Charged particles produced in the target, mainly pions and kaons, are
focused by a pair of toriodal magnets called horns into a \unit[675]{m} 
long decay volume where the mesons decay to muons and neutrinos. The decay region is followed by a hadron absorber where remaining mesons 
and protons are stopped. The neutrino beam then traverses \unit[240]{m} of unexcavated rock before reaching the Near Detector 
located \unit[1.04]{km} from the target.

\begin{figure}
\includegraphics[width=9.25cm]{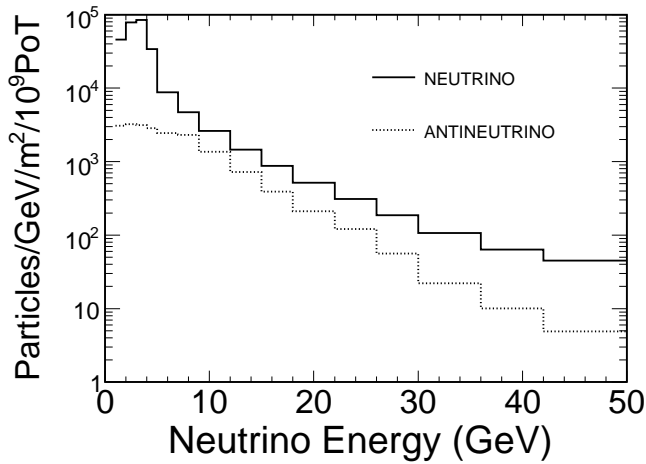}

\caption{\label{fig:gnumi_flux} The muon neutrino and antineutrino flux at the center of the Near Detector as calculated by the NuMI beam simulation.}

\label{beam_spectra}
\end{figure}

Data for this analysis were collected in ``low energy'' beam mode in which the downstream end of the 
target is placed \unit[10]{cm} from the neck of the first focusing horn 
and the current in the horns is \unit[185]{kA}, with the polarity set
to focus positively charged mesons. The Monte Carlo simulation predicts
the composition of the event sample to be
92.9\% $\nu_{\mu}$, 5.8\% $\bar{\nu}_{\mu}$, and 1.3\% $\nu_{e}+\bar{\nu}_{e}$. Fig. \ref{beam_spectra} shows the simulated flux spectrum 
of the $\nu_{\mu}$ and $\bar{\nu}_{\mu}$ in the beam. The $\nu_{\mu}$ component of the beam, which results primarily from 
focused $\pi^+$ and  $K^+$, peaks between 3 and \unit[4]{GeV} with a long tail.
The $\bar{\nu}_{\mu}$ component arises mainly from low transverse momentum 
$\pi^-$ and $K^-$ traveling through the neck of both horns, where they undergo little defocusing. 
This results in a spectrum with no focusing peak and greater mean energy.

\subsection{Near Detector}

The Near Detector is a tracking calorimeter composed of planes
of magnetized iron and plastic scintillator. A toroidal magnetic
field with an average strength of \unit[1.3]{T} provides a measure of muon momentum
from curvature and is used to distinguish $\nu_{\mu}$ and $\bar{\nu}_{\mu}$
CC interactions based on the charge sign of the final state muon. In
normal operational mode the field is set to focus negative muons.

\begin{figure*}[t]
\includegraphics[width=8cm,height=4cm]{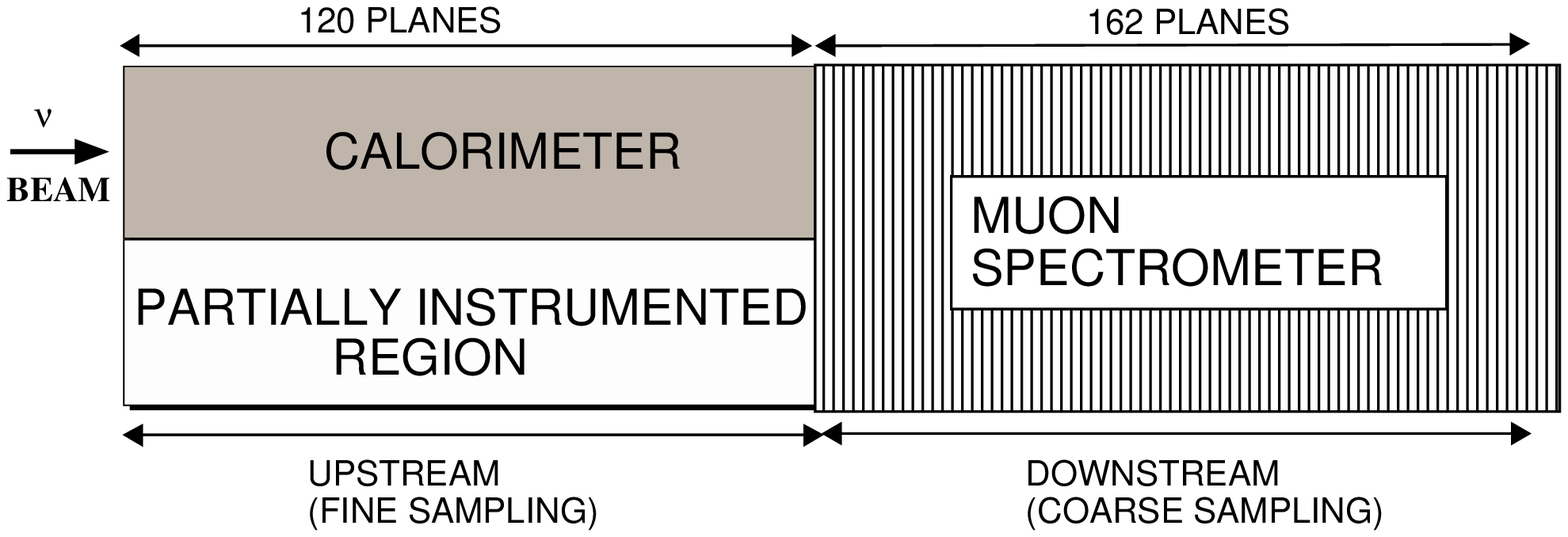}
\hspace{0.1in}
\includegraphics[width=8cm,keepaspectratio=true]{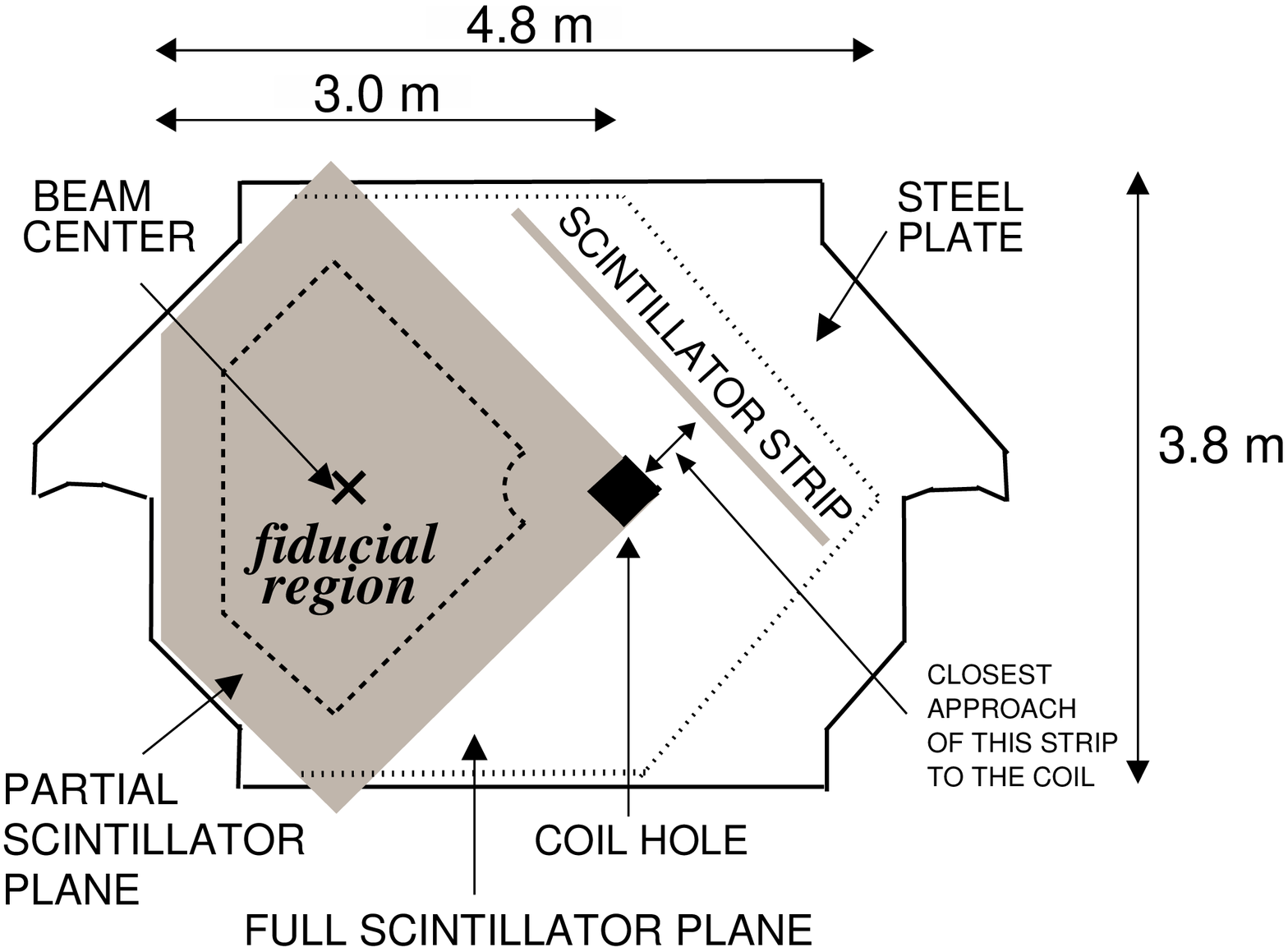}

\caption{ \label{long_view}\label{fig:detector} Left: top view of the Near Detector, showing the calorimeter and muon spectrometer. The drawing is not to scale. Right: transverse view of a Near Detector plane. The shaded area shows a partially instrumented active scintillator plane and the dashed line within shows the boundary of the fiducial region. The dotted line shows the outline of a fully instrumented scintillator plane.}

\end{figure*}

The Near Detector, illustrated in Fig. \ref{long_view}, consists of 282 steel
plates, \unit[2.54]{cm} thick, of which 152 are instrumented with 1~cm thick
scintillator planes. The scintillator planes are made of \unit[4.1]{cm} wide
strips oriented $\pm45^\circ$ with respect to the vertical and
alternating $\pm 90^\circ$ in successive planes. The strips are read
out with wavelength shifting fibers connected to multi-anode
photo-multiplier tubes (PMT). 
Every fifth plane throughout the
detector is fully instrumented with a scintillator layer. 
In the upstream calorimeter region, comprising the first 120 planes, each of the four intervening planes has partial scintillator coverage. 
The calorimeter region is used to measure energy deposited by
neutrino-induced hadronic showers. Event verticies are required to be
within a fiducial volume contained in the calorimeter.
The downstream 162 planes of the detector form the muon spectrometer.

In the low energy NuMI beam configuration, the typical interaction rate in the
Near Detector is about 16 events in a \unit[10]{$\mu$s} spill. Events are
separated using timing and spatial information.
The events accepted for this analysis were from
interactions occurring during a $\unit[13]{\mu s}$ long
gate synchronized to the beam spill. The readout electronics
continuously digitize the PMT signals in \unit{19}{ns} samples without
deadtime throughout the spill.
In between beam spills, cosmic ray muon data are recorded with less than 1\% deadtime.

\label{sub:Calibration}

The detector is calibrated in several steps that convert the raw PMT
signal to deposited energy~\citep{nim_paper}. The non-linearity of the electronics is 
measured with charge injection; relative PMT gains are measured with an 
in-situ light injection system; variations in the light output between 
scintillator strips and along the strips are corrected with cosmic ray muons 
and a radioactive source scanner. Cosmic ray muons which stop in the 
detector are used to calibrate the measured signal to energy lost
by muons passing through the scintillator strips.
The detector simulation is tuned to emulate the actual detector response
at all stages in the calibration chain.

\subsection{Beam and Detector Monte Carlo Simulation \label{sub:Simulation}}

A Monte Carlo simulation is used to model the production
of the neutrino beam, interaction of neutrinos in and around the
detector, and the detector response, which is simulated using {\sc geant3}~\citep{geant3}. 
The beam model includes a simulation of
secondary hadron production from proton interactions~\citep{fluka05}
and the propagation of these hadrons. 
Their reinteraction and decay products 
are also tracked through the target, magnetic horns, and decay region. 
This simulation produces an initial estimate of the flux, which
is later replaced by the flux extracted using the method described below.

Neutrino interactions in the detector are simulated using 
the {\sc neugen3}~\citep{neugen} event generator. 
The simulation of quasi-elastic interactions, which
dominate at low energies, is based on the Llewellyn-Smith~\citep{qel}
model, while intermediate-energy resonance interactions are simulated
according to the Rein-Sehgal model~\citep{res1,res2}.  Both 
models assume a dipole parametrization of the axial part of the cross
section that depends on the axial mass parameters $M_{A}(QEL)$ and
$M_{A}(RES)$, taken to be 0.99$\pm$0.15 and 1.12$\pm$\unit[0.17]{GeV}, 
respectively.  A transition is made between resonance
production and the DIS model by phasing
out the former and phasing in the latter 
over the hadronic invariant mass range, 1.7$<W<$\unit[2.0]{GeV}.  
The sum of the resonance and DIS contributions are constrained
to match total cross section data.

DIS interactions, which dominate at
high energy, are based on an effective leading order model
by Bodek and Yang~\citep{bodek-yang}. The Bjorken scaling
variable $x$ is replaced by an effective scaling variable that depends
on two parameters $A_{ht}$ and $B_{ht}$, where $A_{ht}$ accounts
for target mass effects and higher-twist terms.
$B_{ht}$ depends on the transverse
momentum of the initial state quark. The model is fit to charged
lepton scattering data~\citep{bodek-yang}  and gives the parameters $A_{ht}$ and $B_{ht}$
and correction factors ($C_{v1u}$, $C_{v2u}$, $C_{v1d}$, $C_{v2d}$,
$C_{s1d}$ and \textbf{$C_{s1u}$}) for valence and sea up and down quark parton distribution
functions.  
The uncertainties on these parameters were not readily available
so a study was performed to estimate
them and their effect on this cross section measurement
(see Sec. \ref{sec:Systematics}).

The cross section in the transition region from resonance to DIS is expressed
as a sum of a pure-resonance cross section and a non-resonance contribution
from DIS. The sum is tuned to describe low multiplicity final state data in
this region~\citep{neugen}. For DIS interactions, the final state hadronic 
system is modeled with KNO scaling~\citep{kno}, which transitions to
{\sc pythia}/{\sc jetset}~\citep{Sjostrand:1993yb} at hadronic 
invariant mass W=\unit[3]{GeV}.  The total neutrino cross section is tuned by
a scale factor so that the cross section at \unit[100]{GeV} matches the world
average of measurements.

The dynamics of hadron formation in the target nucleus and reinteraction of hadrons after formation 
modify the visible hadronic shower energy. 
These effects are simulated using a cascade Monte Carlo anchored to $\pi$N, pN and $\pi$Fe and pFe scattering data 
and validated against neutrino-deuterium and neutrino-neon scattering data~\citep{inuke,inuke_data}. A treatment
of hadron formation time is included~\cite{ammosov}.

\section{analysis\label{sec:analysis}}

The CC total cross sections are measured from the 
inclusive CC scattering rate, \gcc{}, and the incident neutrino flux, \flux{}. 
A sample of CC events, \ncc{}, is selected and 
then corrected for acceptance and backgrounds to determine \gcc{}.  
A flux sample, \fsam{}, consisting of the subset of \ncc{} 
with low $\nu$, (in the lab frame $\nu=E_{had}$, the energy measured at the hadronic 
vertex),
is also defined and corrected for acceptance, backgrounds, and for  
a small energy dependence using our Monte Carlo model to yield \flux{}.
The event reconstruction and selection of these samples to form the 
cross section are described in this section.

The data used in this analysis were collected between June 2005 and April 2007 and correspond to an exposure of 2.45$\times10^{20}$ PoT. The MC sample is almost double the data, corresponding to 4.4$\times10^{20}$ PoT.

\subsection{Event Reconstruction}

Neutrino events are identified  using the timing and spatial pattern of energy deposited in the scintillator strips. Muon tracks are recognized as a string of 
hit strips typically spanning more than 10 steel plates. For muons that stop in the detector the energy is computed from range according to the energy loss tables of Groom, {\it et al.}~\citep{range}. 
 A systematic uncertainty of 2\% is assigned to the energy measured from range, 
arising from uncertainties in the range tables, the variation in material composition and the accuracy of our track length reconstruction. 
The momentum of muons exiting the detector are measured using the curvature of their trajectory in the detector's magnetic field.  
A 4\% systematic uncertainty is assigned to our knowledge of the absolute muon momentum measurement from curvature. This is 
assessed by comparing the energy measured with curvature to the independent measurement from range using tracks 
that stop in the detector, and by folding in underlying uncertainties in the detector's magnetic field~\cite{rustem_thesis}. 
The resolution for muon momentum measured from range is 5\% while that measured from curvature has
non-Gaussian tails and width of approximately 10\%.

The vertex of a neutrino interaction is taken to be at the start of a reconstructed track.
Hit strips near the vertex which are not included in the track are identified as coming from hadrons produced 
in the interaction. Their summed signal is converted to energy using a lookup table derived from simulated showers
to form the hadronic shower energy, $E_{had}$. The response of our detector to single hadrons 
was measured in an exposure of a smaller version of the detector to a test beam~\citep{caldet}. 
The measured test beam detector response was used to tune our simulations. The absolute energy scale of the detector's
response to hadronic particles is modeled to an accuracy 
of 5.6\%~\cite{kordosky_thesis,prd_1e20}, which we take as the hadronic
energy scale uncertainty in the cross section measurement (see Sec. \ref{sec:Systematics}).

\subsection{CC Event Selection\label{sub:Event-Selection}}

The inclusive charged-current sample \ncc{} is selected using the following criteria:

\begin{enumerate}

\item \emph{Fiducial volume} - Selected events have a vertex position along the detector axis between 0.5 and \unit[4.0]{m}, measured from the upstream 
face of the detector. In the plane transverse to the detector axis, the vertex is required to be more than \unit[0.5]{m} from the edge of an 
active scintillator plane and outside of a \unit[0.8]{m} radius centered at the coil hole.
The outline of the fiducial region is shown in Fig. \ref{fig:detector}.

\item \emph{Coil hole} - The coil hole is uninstrumented and variations in the material composition and magnetic field are somewhat larger in the region around it. To reduce the effect of these uncertainties, events with tracks that spend a significant fraction of their path-length near the hole are removed from the event sample. 
A minimum of 95\% of hit strips in the event are required to be 
further than \unit[0.3]{m} from the center at closest approach (see Fig. \ref{fig:detector}). 

\item \emph{Track Energy} - The energy of the muon must be greater than \unit[1.5]{GeV}.  This requirement rejects neutral-current (NC) background events, which populate the low energy region, and short, poorly reconstructed tracks. 
\item \emph{Track Quality} - The track fitting procedure yields a measurement of the muon momentum with an associated uncertainty.
The track fit is required to be convergent and have an uncertainty of less than 30\%.  
In addition, we require the track's longitudinal start positions in each view to be less than six planes apart.

\item \emph{Neutrino Energy} - The reconstructed neutrino energy, $E$, which is the sum of the track and shower energies, is required to be greater 
than \unit[3]{GeV} (\unit[5]{GeV}) for the neutrino (antineutrino) sample and less than  \unit[50]{GeV}.
The minimum energy requirements are imposed to minimize the overlap of the inclusive CC sample and the flux sample, which is substantial
below these values. Above the maximum energy cut, resolution of the track momentum measurement from curvature degrades as the tracks become 
straighter.

\end{enumerate}

The event sample is divided into two categories depending on whether
the track stops in or exits the detector. For exiting events, 
the muon leaves the detector through the back or side, or passes into the
uninstrumented coil hole region.
The stopping and exiting samples
are further differentiated based on whether they end in the upstream
or downstream region (see Fig. \ref{long_view}) because of the difference in sampling
in the two regions.

The $\bar{\nu}_{\mu}$ CC sample is selected by requiring the sign of the track curvature measurement to be positive.
This sample has a higher fractional contamination
from wrong-sign events (misidentified $\mu^-$ tracks) 
due to the much larger $\nu_{\mu}$ component
of the beam. The following additional requirements are imposed to the $\bar{\nu}_{\mu}$ CC
sample to reduce this contamination:

\begin{enumerate}
\item \emph{Bend away from coil} -
We require that the track 
bend  away from the magnet coil hole to
reject positive charge track candidates whose curvature 
is mismeasured by the tracker.
An angle is defined in the transverse plane by forming 
a straight line from the extrapolated track end point in absence of a magnetic field to 
the observed track end point, and a line from the magnet
coil hole center to the observed interaction point~\cite{rustem_thesis}.  
For a particle bending
toward the coil, this angle will be near $\pi$ radians, while for
a defocused $\mu^+$ it will be near 0 or 2$\pi$ radians.  For the antineutrino
sample, we select a value for this angle less than 1.04 radians
or greater than 5.24 radians.

\item \emph{Number of hit planes} - We keep events in which the difference
in the number of hit planes along the track between the two views is less
than five. Events which are rejected by this cut usually enter
the uninstrumented region in one view yielding an unreliable determination
of the charge sign.
\item \emph{Downstream exiting tracks} - Only events with tracks
that exit the detector in the downstream region 
are used for the antineutrino analysis. The
rejected samples have high contamination from NC and 
misidentified $\nu_{\mu}$ CC (wrong-sign) events.
\end{enumerate}
\begin{table*}
\begin{tabular}{|c|c|c|}
\hline 
Selection Criterion & Track Charge $<$ 0 & Track Charge $>$ 0 \tabularnewline
 &  (\%removed) & (\%removed)\tabularnewline
\hline
\hline 
Track vertex in fiducial volume & 3608572 & 841986\tabularnewline
\hline 
$E_{\mu}>$\unit[1.5]{GeV} & 2571917 (28.7\%) & 344110 (59.1\%)\tabularnewline
\hline 
Track Quality Cut & 2351328 (\hspace{0.5em}8.6\%) & 282657 (17.8\%)\tabularnewline
\hline 
$3<E_{\nu}<$\unit[50]{GeV} & 1941019 (17.5\%) & \tabularnewline
\hline 
($5<E_{\nu}<$\unit[50]{GeV} for Track Charge$>$0) &  & 235024 (16.9\%) \tabularnewline
\hline 
Additional $\bar{\nu}_{\mu}$ cuts & - & 159880 (32\%)\tabularnewline
\hline
\end{tabular}

\caption{Effect of the selection criteria on the negative (left) and positive (right)
charge reconstructed track samples. Each row shows the number of events remaining after
each successive cut. The numbers in parentheses show the percentage
of events removed by each cut compared with the previous row.}

\label{tab:cuts_table}
\end{table*}

Table~\ref{tab:cuts_table} shows the effect of the selection criteria on the neutrino and the antineutrino reconstructed samples. The minimum track energy cut has the largest effect, resulting in an approximately 30\% loss in the $\nu_{\mu}$ sample and 60\% in the $\bar{\nu}_{\mu}$ sample. The cut removes primarily 
NC events, which arise from both neutrinos and antineutrinos, and therefore
affects the smaller $\bar{\nu}_{\mu}$ sample more. 
The track quality cut also has a larger effect on the  $\bar{\nu}_{\mu}$ sample. 
A large fraction of the  $\nu_{\mu}$ CC tracks whose charge has been mismeasured due to poor curvature determination
are removed from the   $\bar{\nu}_{\mu}$  sample  by this cut.
After all selections 
have been applied, the inclusive event sample \ncc{} consists of  
$1.94\times10^{6}$ $\nu_\mu$ and $1.59\times10^{5}$ $\bar{\nu}_{\mu}$ events. 

\begin{figure}
\includegraphics[width=9.25cm]{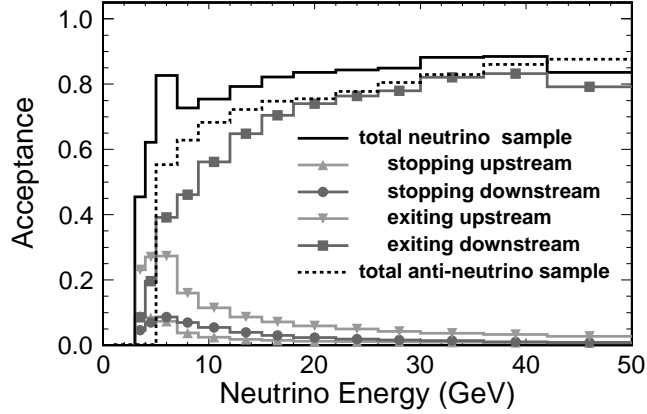}

\caption{The detector acceptance \acc{} for the inclusive charged-current $\nu_\mu$ and $\bar{\nu}_\mu$ samples. Only exiting downstream events are used in the $\bar{\nu}_\mu$ analysis.}

\label{fig:accep_corr}
\end{figure}

The CC sample is organized into energy bins and corrected for detector acceptance \acc{} and backgrounds \bcc{} to obtain 
the CC scattering rate, $\gcc{}=(\ncc{}-\bcc{}) / \acc{}$, where  
$\acc{}$, shown in  Fig. \ref{fig:accep_corr}, represents the number of Monte Carlo events reconstructed in a given bin
divided by the number generated in that bin.
The decrease in acceptance at low energy is due to the minimum muon energy requirement. 
For neutrinos, the shape below \unit[10]{GeV} is determined by the geometry of the detector and overlap of the stopping and exiting samples, which have different resolutions. The contributions from each subsample are also shown in Fig. \ref{fig:accep_corr}. 

\begin{figure*}
\includegraphics[width=8cm]{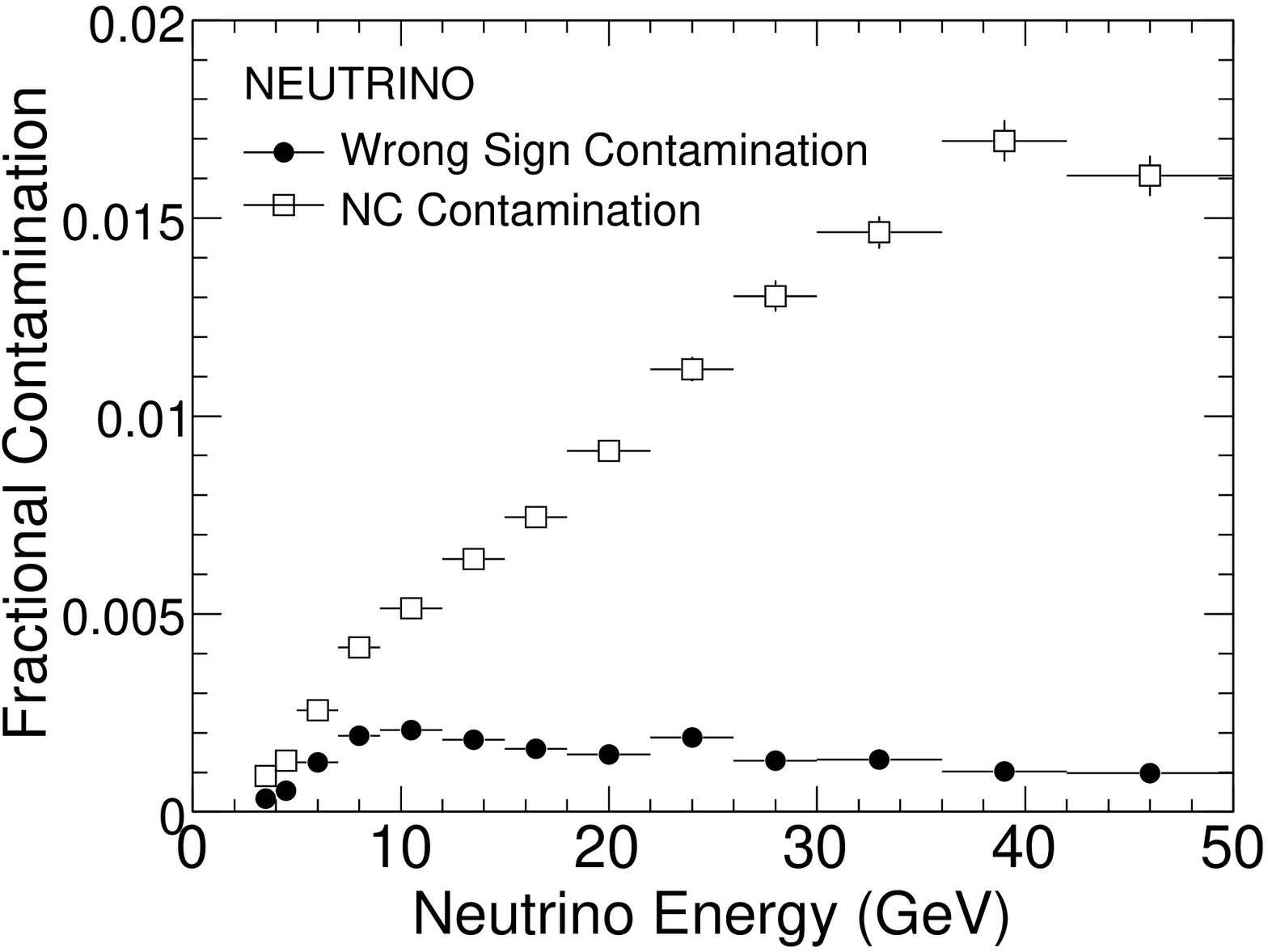}
\includegraphics[width=8cm]{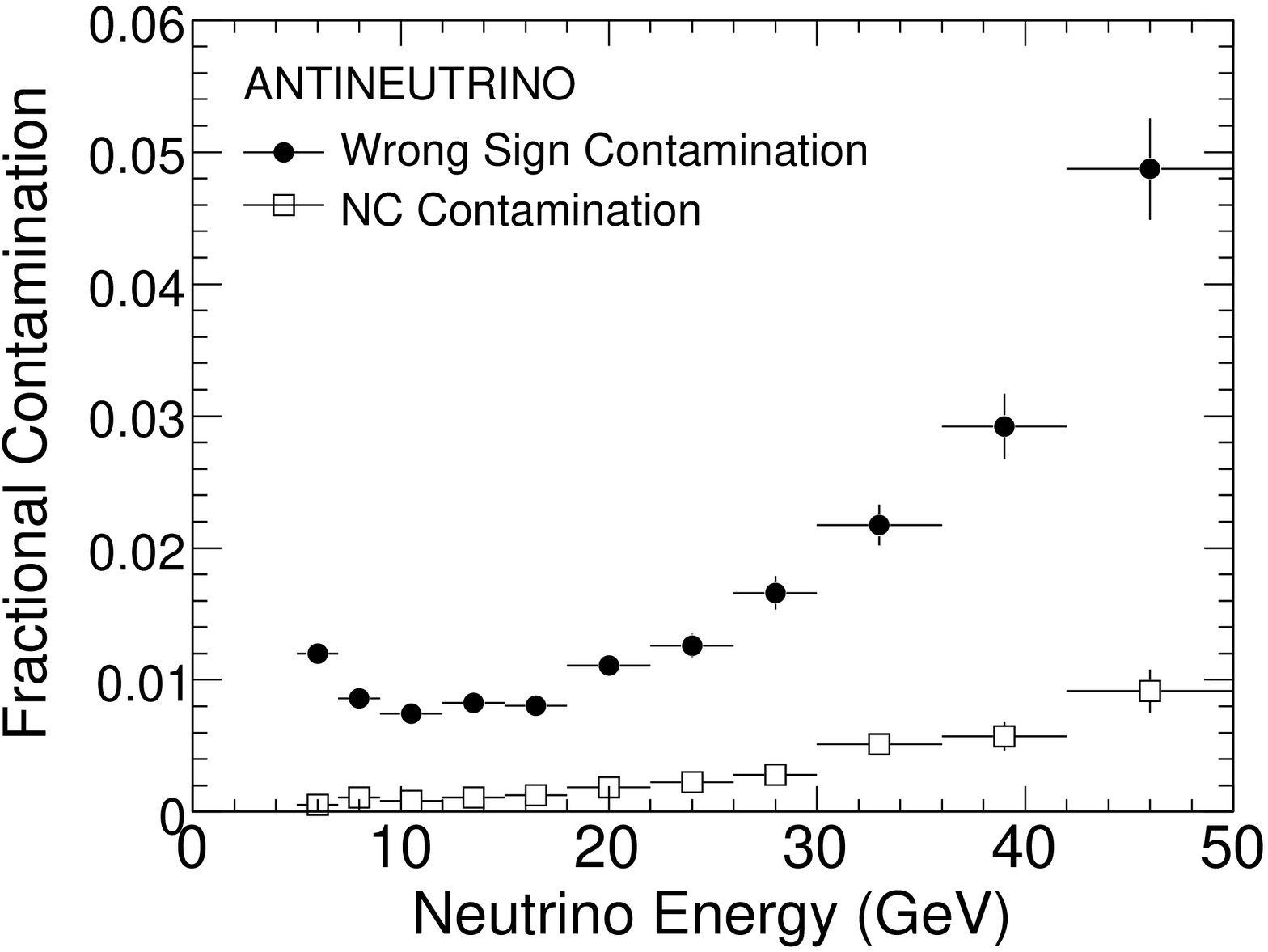}

\caption{Neutral-current and wrong-sign backgrounds in the neutrino (left) and
antineutrino (right) selected charged-current sample as calculated from
the Monte Carlo event sample. The error bars show the statistical
uncertainty only.}

\label{fig:nu_contam}
\end{figure*}

We use our simulation to estimate the backgrounds from NC and wrong-sign events. As shown in Fig. \ref{fig:nu_contam}, the NC background is less than 2\% for both neutrinos and antineutrinos.
It increases with energy due to the contribution from high-inelasticity events in which the primary track is misidentified or perturbed by hits from the hadronic shower particles. The wrong-sign contamination is negligible 
in the neutrino sample but sizable in the antineutrino sample, up
to 5\% at high energy. Wrong-sign background events at low energy come from
$\nu_{\mu}$ CC events in the peak of the neutrino beam, while at higher energies,
the through-going muons have increasingly larger bend
radii, making their charge determination less certain.

\subsection{Flux Extraction \label{sub:Flux-Extraction}}

The ``low-$\nu$'' method~\citep{Mishra:1990ax,seligman} relies on the independence of the
differential cross section, $\frac{d \sigma^{\nu,\nub}}{d\nu}$, with energy in the limit $\nu\rightarrow 0$.
The differential dependence of the neutrino (antineutrino)
cross section, $\frac{d^2\sigma^{\nu,\nub}}{dxdy}$, on inelasticity, $y=\nu/E$, and the Bjorken scaling variable, $x$,
can be written as
\begin{eqnarray}
\label{eqn:disxsec}
\frac{d^2\sigma^{\nu(\overline{\nu})}}{dxdy} &=& \frac{G_F^2 M
 E}{\pi} {\Big(}\Big[1-y(1+\frac{Mx}{2E})
+\frac{y^2}{2}\Big(\frac{1+(\frac{2Mx}{Q})^2}{1+R_L}\Big)\Big] F_2 
\pm \Big[y-\frac{y^2}{2}\Big]xF_3{\Big)}
\end{eqnarray}
where $G_F$ is the Fermi weak coupling constant, $M$ is the
proton mass and $E$ is the incident neutrino energy.
The plus sign in front of the $xF_3$ term is for neutrinos and the minus is for
antineutrinos.
The structure functions $F_2(x,Q^2)$, $xF_3(x,Q^2)$ and  $R_L(x,Q^2)$
depend on $x$ and $Q^2$,
$R_L$ is the ratio of the cross section for scattering
from longitudinally to transversely polarized W-bosons.
For quasi-elastic interactions the cross section can be
written in this form with combinations of form factors replacing the 
structure functions.

Integrating over $x$, the differential dependence on $\nu$ can be written in the
simplified form
\begin{equation}
\frac{d\sigma^{\nu,\nub}}{d\nu}=A\left(1+\frac{B}{A}\frac{\nu}{E}-\frac{C}{A}\frac{\nu^2}{2E^2}\right) \label{eq:dsigmadnu}.
\end{equation}
The coefficients $A,$ $B$, and $C$ depend on integrals over
structure functions,

\begin{eqnarray}
A &=& \frac{G^2_FM}{\pi}\displaystyle\int F_2(x) dx,\\
\nonumber
B &=& -\frac{G^2_FM}{\pi}\displaystyle\int \Big( F_2(x)  \mp xF_3(x)\Big) dx,\\
\nonumber
C &=& B - \frac{G^2_FM}{\pi}\displaystyle\int F_2(x) \tilde{R} dx,
\nonumber
\end{eqnarray}
where
\begin{eqnarray*}
\tilde{R} &=& \left(\frac{1+\frac{2Mx}{\nu}}{1+R_{L}}-\frac{Mx}{\nu}-1\right).
\end{eqnarray*}
The factor A is nearly
the same for neutrino and antineutrino probes\footnote{For an isoscalar target with only $u$ and $d$ quarks $F_2^{\nu}=F_2^{\bar{\nu}}$, assuming
isospin symmetry. Including $s$ quarks and CKM mixing gives a small difference term $F_2^{\nu}-F_2^{\bar{\nu}}=-\frac{1}{2} V_{us}^2 (u_v+d_v)$, 
where $u_v$ and $d_v$ are the valence quark
distributions. We apply a correction to account for this term to the antineutrino normalization.}
, however,
the magnitude of the coefficient  $B$ 
is larger for antineutrinos, where the $xF_3$ contribution is added, compared with the
neutrino case where the term is subtracted.
As discussed later, this makes the energy dependence correction needed in this
method larger for 
the antineutrino flux shape. The $C$ term, which depends on $R_L$, is small.


For small ${\nu}/{E}$, Eq. \ref{eq:dsigmadnu} shows that the differential cross section becomes 
independent of energy and is equal to the same constant, $A$, for neutrinos and antineutrinos.
Multiplying both sides 
by the flux, $\phi(E)$, and taking the limit $\nu\rightarrow0$
gives 
\begin{equation}
\frac{dN}{d\nu}\Big|_{\nu\rightarrow0}=A\Phi(E).
\end{equation} 
Therefore, the flux in a given energy bin can be approximated using the number of events at low $\nu$.

We account for the small ${\nu}/{E}$ and $({\nu}/{E})^2$ 
dependence resulting from a finite $\nu_0$
in Eq. \ref{eq:dsigmadnu} using a ``low-$\nu$'' correction 
\begin{equation}
\label{eq:nucor}
\nucor = \frac{\sigma(\nu<\nu_{0},E)}{\sigma(\nu<\nu_{0},E\rightarrow\infty)}
\end{equation} 
that is calculated from our cross section model.
The term  $\sigma(\nu<\nu_{0},E)$ is the value of the integrated cross section below our chosen $\nu_0$ cut at energy E,
and $\sigma(\nu<\nu_{0},E\rightarrow\infty)$ is its value in the high energy limit.

This correction is applied to our selected 
flux sample, \fsam{}, consisting of the 
subset of \ncc{} with $\nu < \nu_0$ that is subsequently corrected for acceptance, \aflux{}, and backgrounds, \bflux{},
\begin{equation}
\phi^{\nu(\overline{\nu})}(E)= \frac{\fsam{}-  \bflux{}}{  \nucor \times \aflux{}}.
\end{equation}
This yields the shape of the flux with energy. A normalization factor $H^\nu$, determined using external data,
must be applied to give the absolute flux,
$\Phi^{\nu(\overline{\nu})}(E)= H^\nu\phi^{\nu(\overline{\nu})}(E)$,
as described in the next section.

\begin{figure*}
\includegraphics[width=8cm]{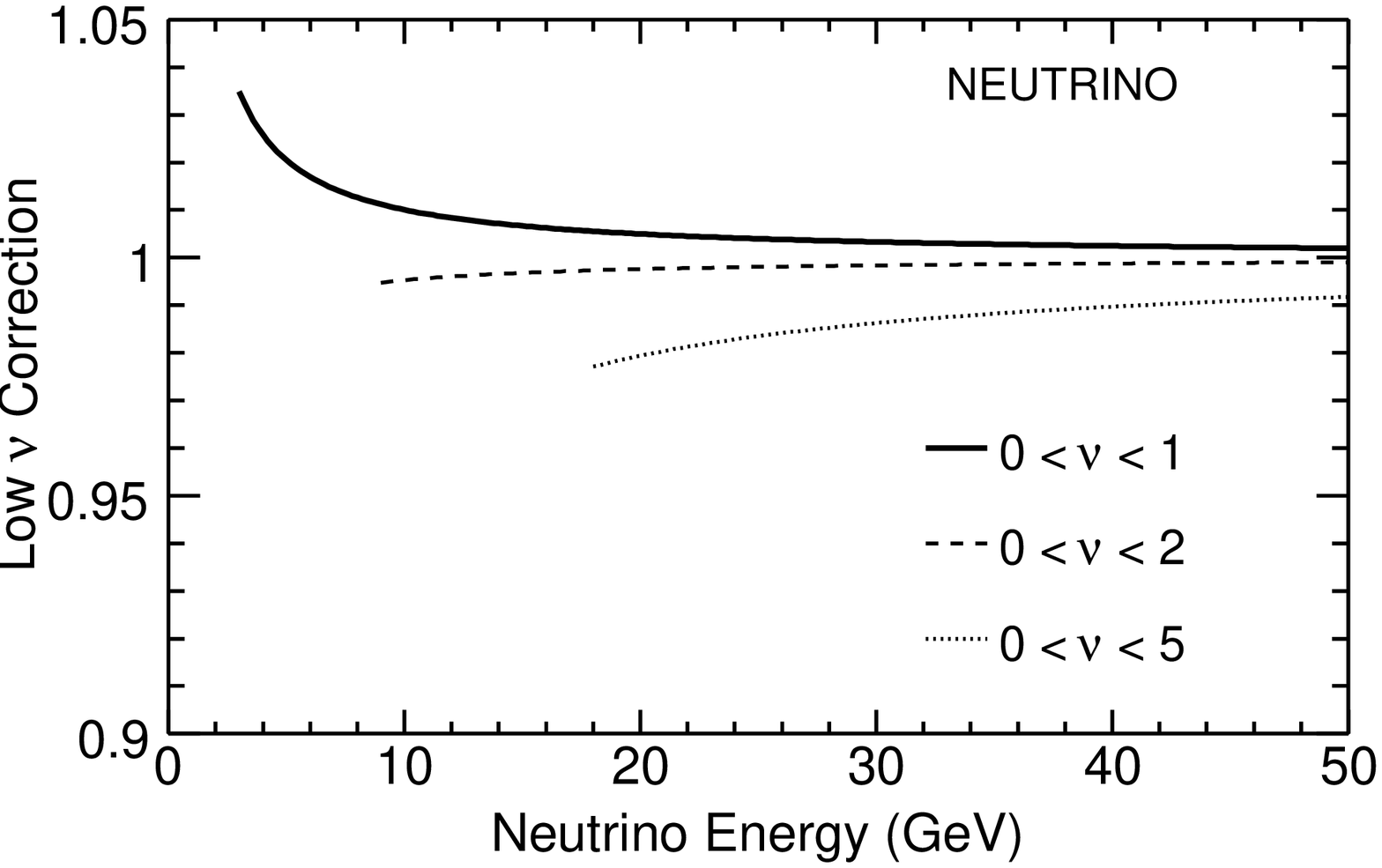}
\includegraphics[width=8cm]{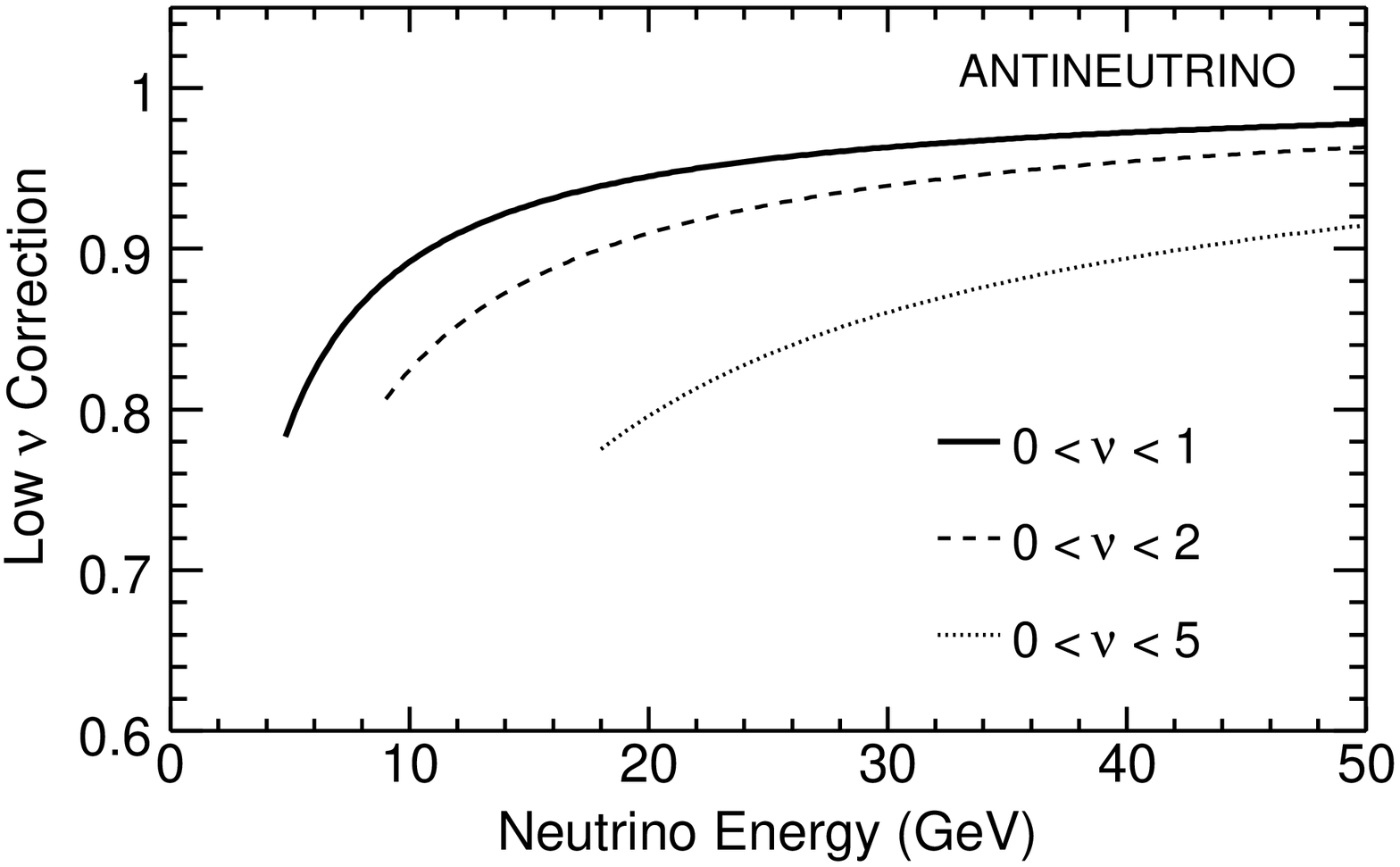}

\caption{The low-$\nu$ correction, \nucor{}, applied to the flux sample for neutrinos (left) and antineutrinos (right).
The solid line shows \nucor{} for $\nu<$\unit[1]{GeV} applied from 3-\unit[9]{GeV}
(5-\unit[9]{GeV} for $\overline{\nu}$), dashed line 
for $\nu<$\unit[2]{GeV} applied from 9-\unit[18]{GeV}, and 
the dotted line for $\nu<$\unit[5]{GeV} applied above \unit[18]{GeV}.}
\label{fig:lownu}
\end{figure*}

Our choice of $\nu_0$ trades statistical precision for modeling uncertainty in determining $\nucor$.
To improve the statistical precision, we increase $\nu_{0}$ with energy while keeping the ratio $\nu/E$ and the resulting 
model dependence small. We set $\nu_{0}=$\unit[1]{GeV} for events with $E_{\nu}<$\unit[9]{GeV}, $\nu_{0}=$\unit[2]{GeV} for 9$<E_{\nu}<$\unit[18]{GeV} and $\nu_{0}=$\unit[5]{GeV} for $E_{\nu}>$\unit[18]{GeV}. Fig. \ref{fig:lownu} shows the size of the low-$\nu$ correction for neutrino and antineutrino samples. The correction for neutrinos is about 3\% at \unit[3]{GeV} and for antineutrino is about 20\% at \unit[5]{GeV}. 

The stronger inelasticity dependence of the antineutrino cross section results in the much larger correction for antineutrinos. 
In addition, antineutrino CC interactions have lower inelasticity on average, which causes a large 
overlap between the cross section and the flux samples.
The overlap decreases with energy from 90\% at \unit[3]{GeV} to about 60\% at \unit[6]{GeV} for antineutrinos,
whereas for neutrinos it is  60\% at \unit[3]{GeV} and below 30\% above  \unit[6]{GeV}~\cite{debdatta_thesis}. 
We therefore  restrict our analysis to the region above \unit[5]{GeV} for the antineutrino sample.
 
The low-$\nu$ correction introduces a model dependence and model uncertainty to the flux determination. We account for this uncertainty in the flux by varying the model parameters described in Sec.~\ref{sub:Simulation} and re-calculating the flux. The change in the correction when the model is varied is 1\% or less because it is a fractional term in which the numerator and denominator are similarly affected.

\begin{figure}
\includegraphics[width=9.25cm,height=7cm]{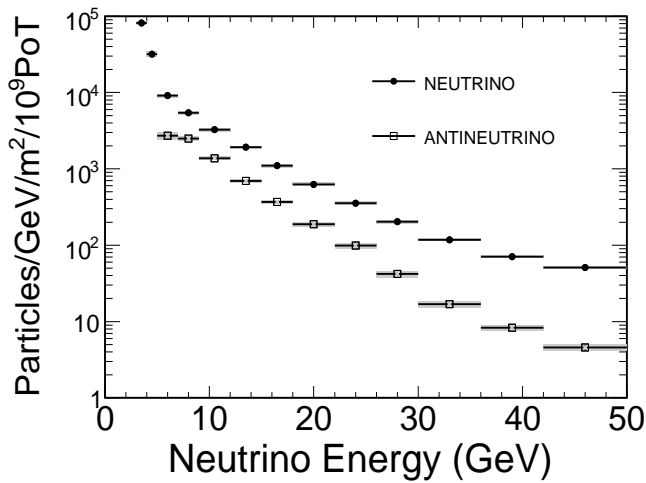}

\caption{Extracted flux as a function of neutrino energy. 
The error bars are too small to see on this log-scale plot,
but are given in Table~\ref{tab:flux_table}. }

\label{fig:extracted_flux}
\end{figure}

Fig. \ref{fig:extracted_flux} and Table~\ref{tab:flux_table} show
the extracted neutrino and antineutrino fluxes 
in the selected fiducial volume after normalization.
The systematic uncertainties on the extracted
flux are discussed in Sec.~\ref{sec:Systematics}. 
We correct the input flux model shown in Fig. \ref{beam_spectra} by reweighting
the simulation with the ratio of the extracted flux to the original simulated 
flux.  The resulting corrections to the initial simulated flux are consistent
with those obtained by a different technique~\cite{prd_1e20} used for the 
MINOS oscillation analyses.

\begin{table}
\begin{tabular}{|c|c|c|c|c|}
\hline 
~~~~$E$ bin~~~~ & ~~~~$\nu$ Flux~~~~ & ~~~~Error~~~~ & ~~~~$\bar{\nu}$ Flux~~~~ & ~~~~Error~~~~\tabularnewline
\hline
\hline 
\multicolumn{1}{|c|}{(GeV)} & \multicolumn{4}{c|}{Particles/GeV/$m^{2}$/$10^{9}$ PoT}\tabularnewline
\hline 
3-4 & 8.05$\times 10^4$  & 5.2$\times 10^3$  & -  & - \tabularnewline
\hline 
4-5 &  3.06$\times 10^4$    &   2.4$\times 10^3$  & - & - \tabularnewline
\hline 
5-7 & 9.07$\times 10^3$   &5.3$\times 10^2$   & 2.80$\times 10^3$  & 330\tabularnewline
\hline 
7-9 & 5.18$\times 10^3$   &  3.5$\times 10^2$ & 2.32$\times 10^3$  & 170\tabularnewline
\hline 
9-12 &  3.21$\times 10^3$   & 2.2$\times 10^2$ & 1.32$\times 10^3$  & 85\tabularnewline
\hline 
12-15 & 1.94$\times 10^3$  &1.0$\times 10^2$  & 6.89$\times 10^2$  & 42\tabularnewline
\hline 
15-18 &  1.09$\times 10^3$  & 65 & 3.79$\times 10^2$ & 24\tabularnewline
\hline 
18-22 & 629 & 37  & 190 & 14\tabularnewline
\hline 
22-26 & 348   & 20  & 86.3 & 7.8\tabularnewline
\hline 
26-30 & 200  & 13 & 40.1 & 3.9\tabularnewline
\hline 
30-36 & 119 & 6.8 & 19.3 & 1.9\tabularnewline
\hline 
36-42 & 72.2  & 3.9 & 9.6 & 0.9\tabularnewline
\hline 
42-50 & 51.6 & 2.8 & 4.9 & 0.5\tabularnewline
\hline
\end{tabular}

\caption{Measured flux as a function of neutrino energy. Statistical, systematic and normalization uncertainties are included in the error estimate.}

\label{tab:flux_table}
\end{table}

\begin{figure*}
\includegraphics[width=8cm]{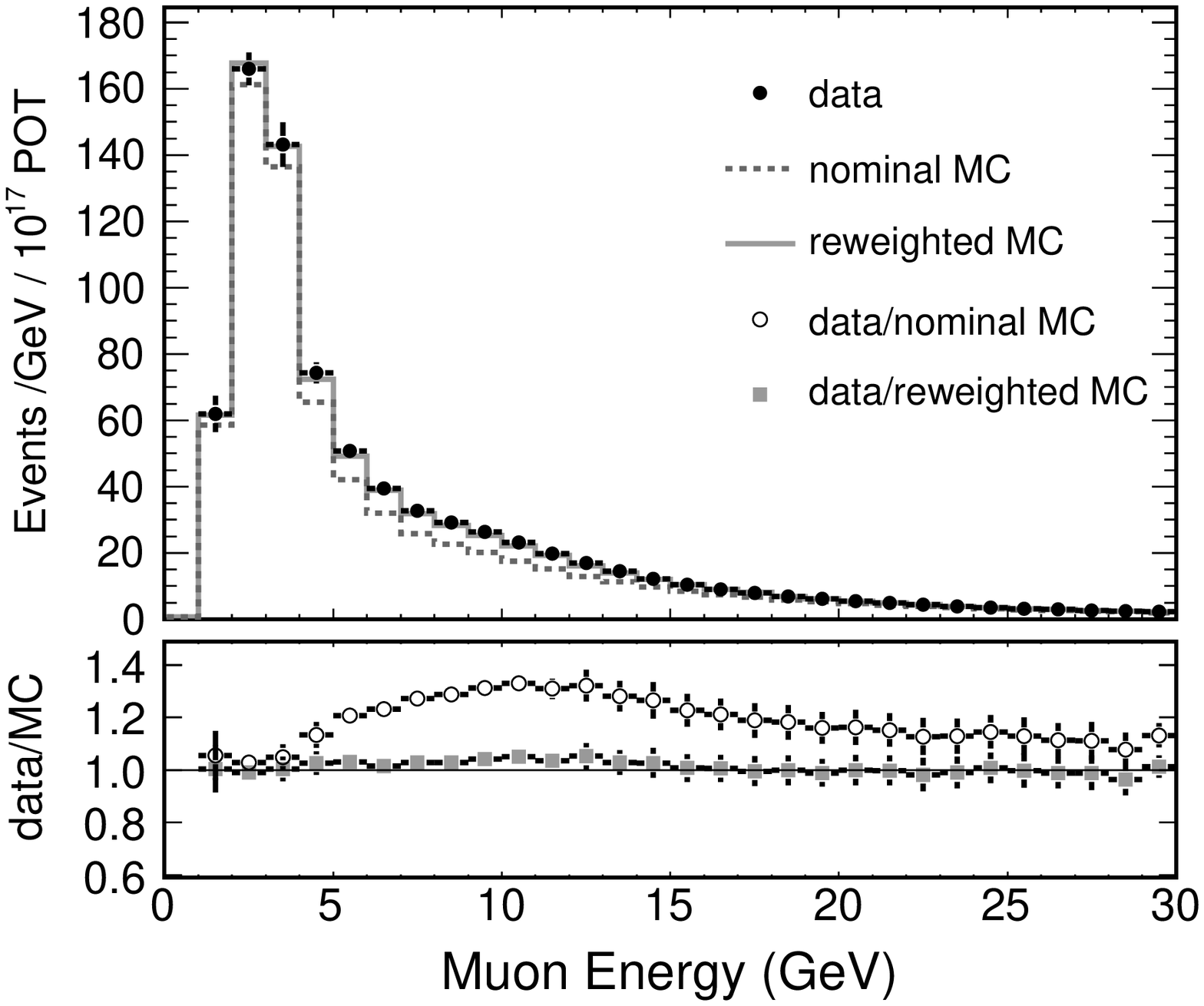}
\includegraphics[width=8cm]{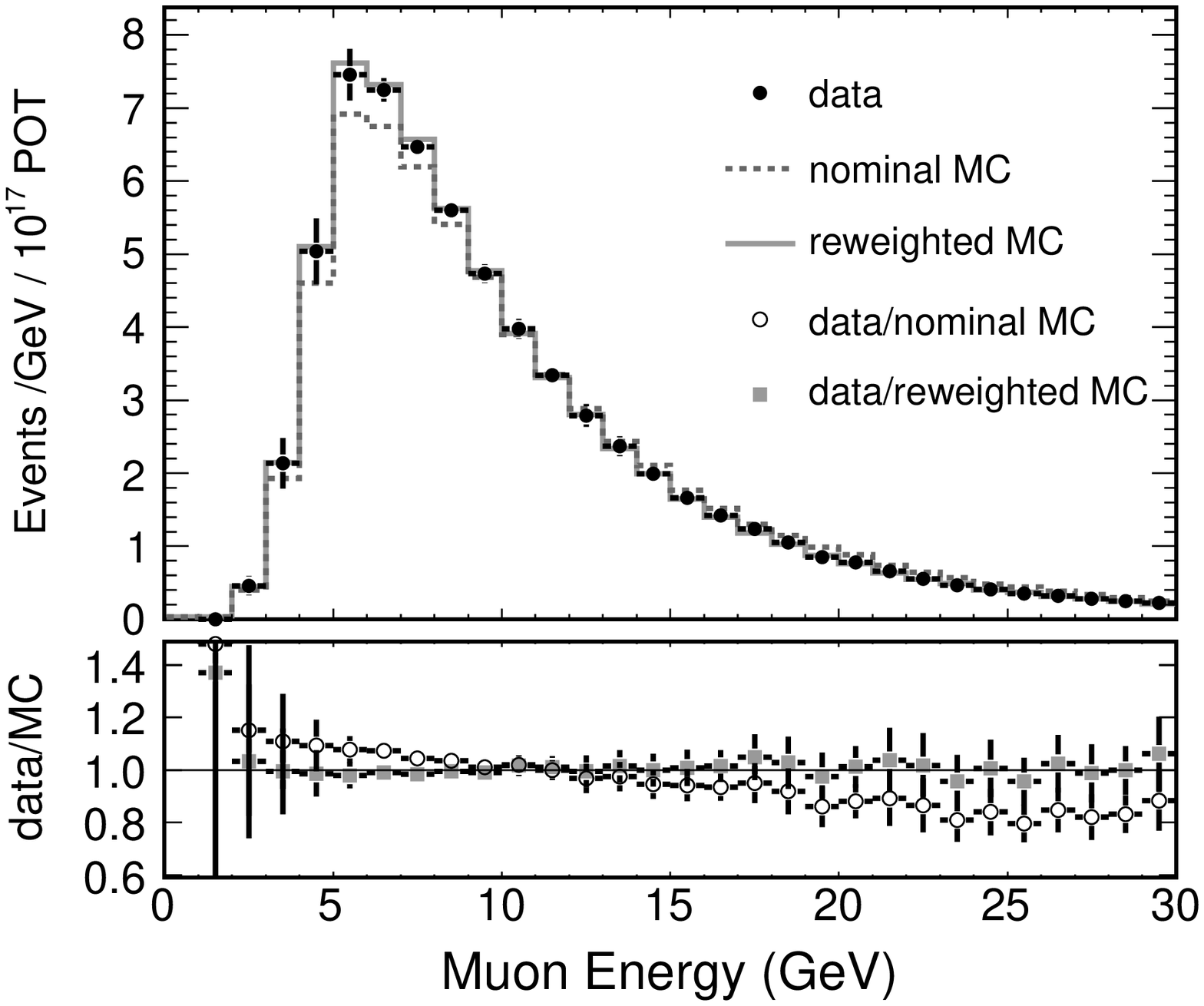}

\includegraphics[width=8cm]{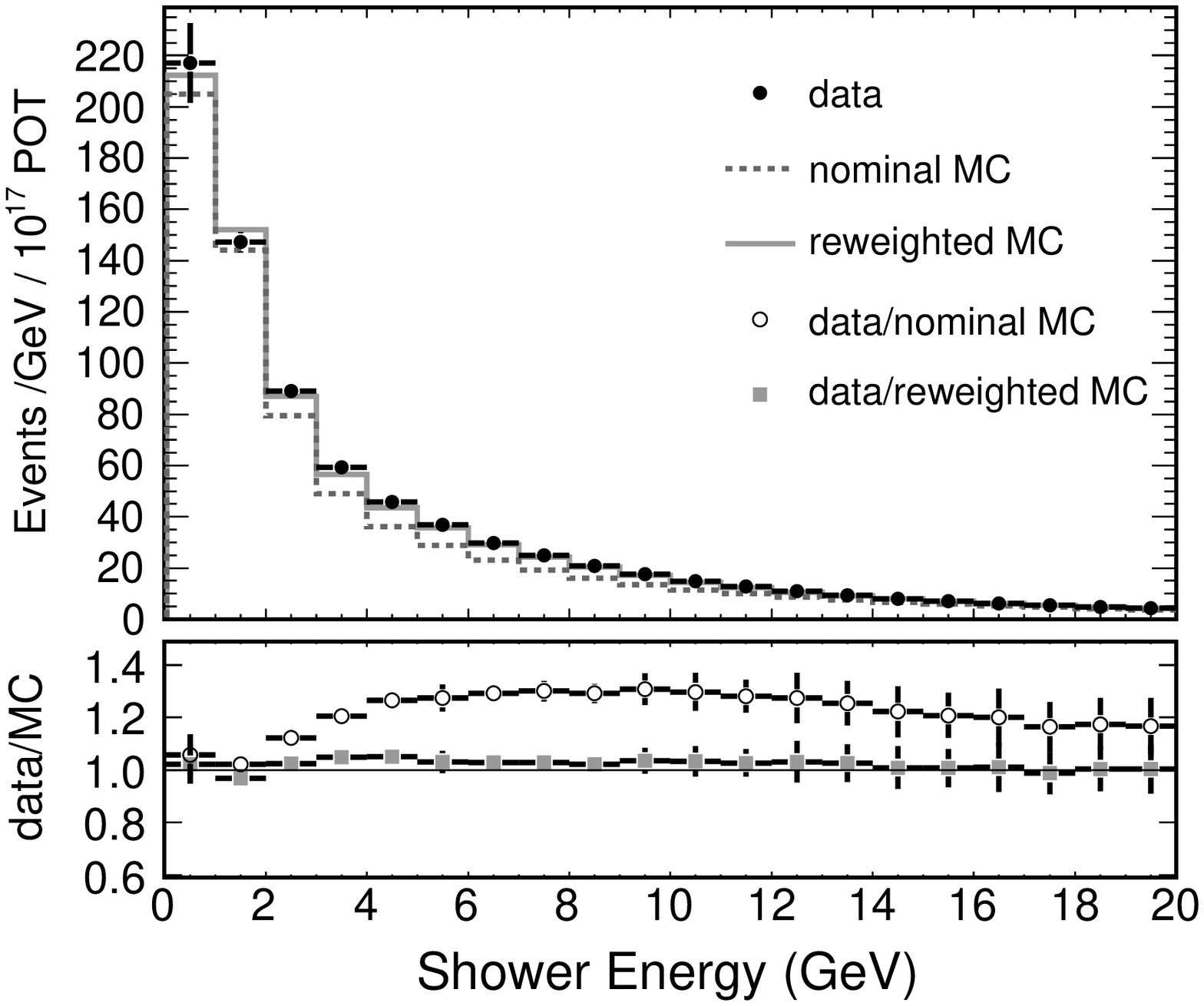}
\includegraphics[width=8cm]{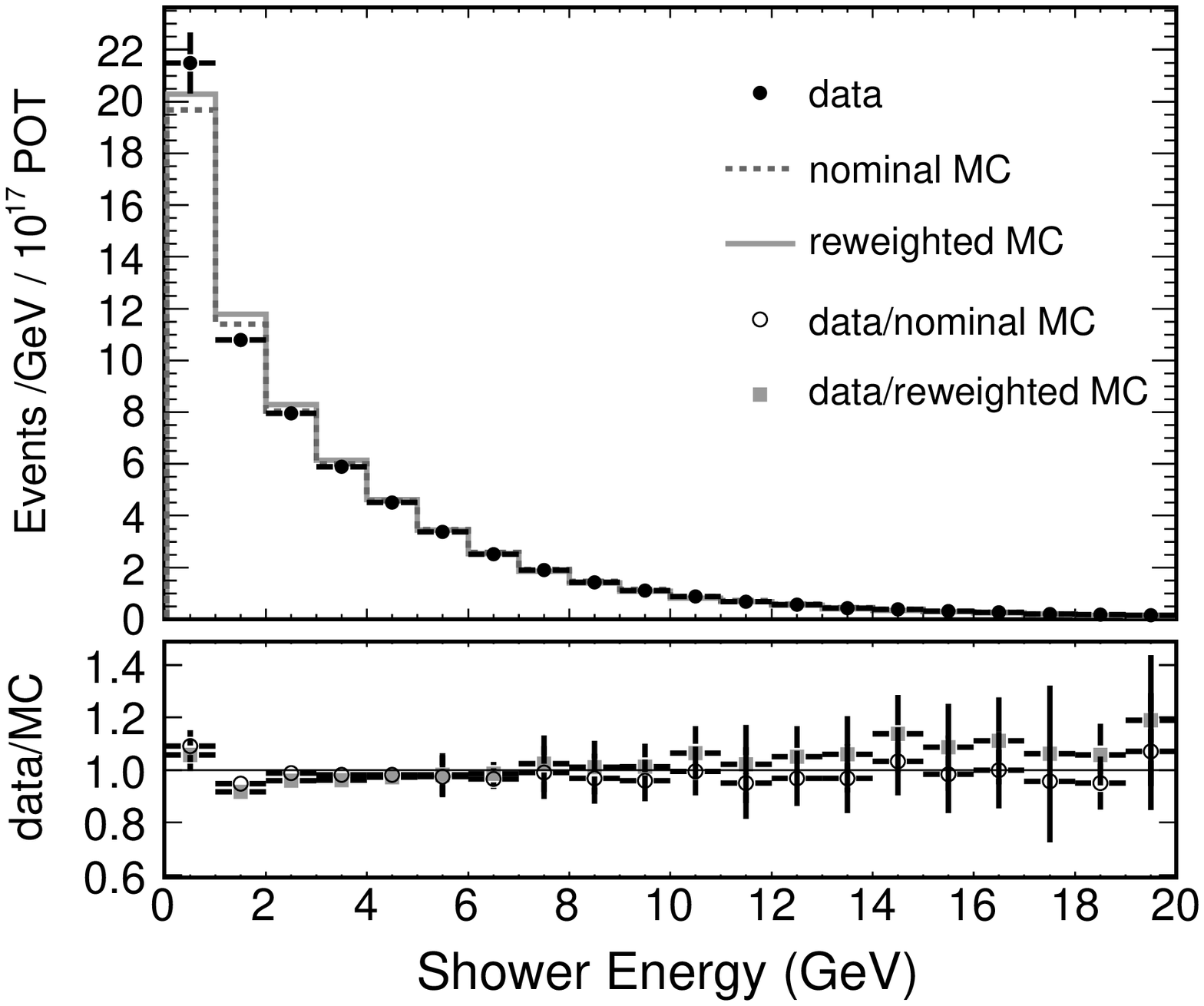}

\includegraphics[width=8cm]{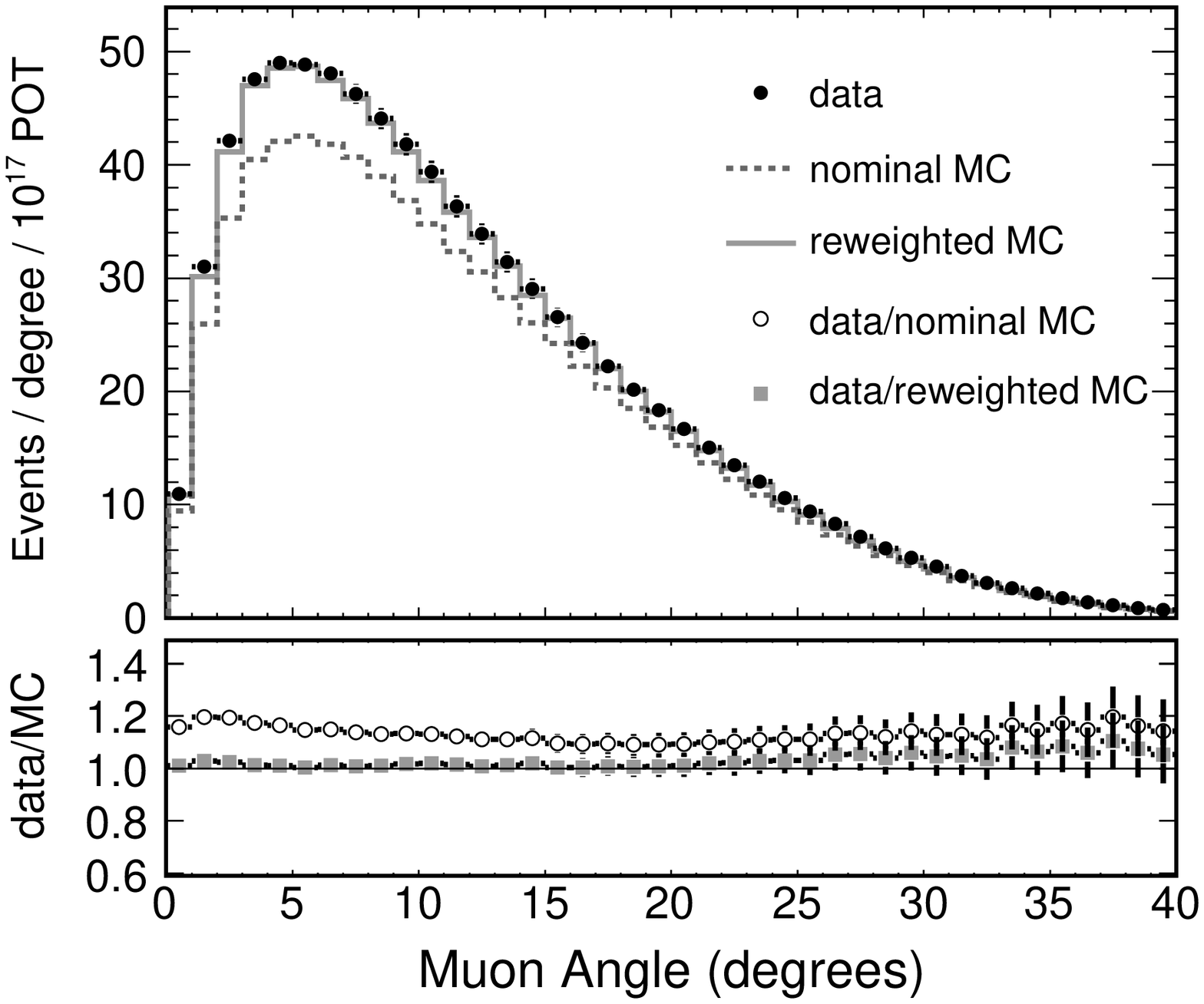}
\includegraphics[width=8cm]{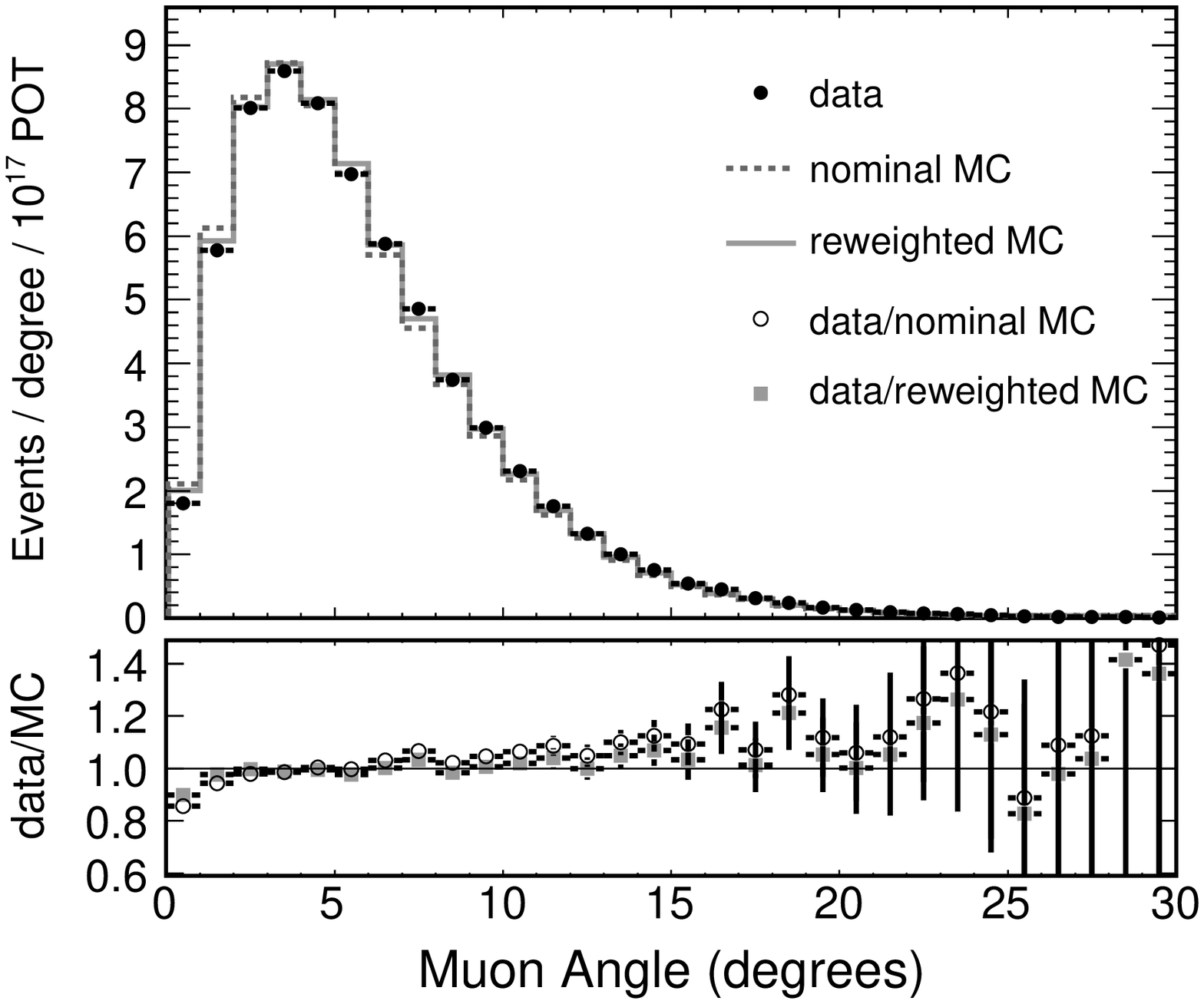}

\caption{Comparison of data and Monte Carlo simulation distributions for the
kinematic variables $E_{\mu}$, $E_{had}$, and $\theta_{\mu}$ for
neutrinos (left) and antineutrinos (right). The points show the data, the dashed
line shows the nominal Monte Carlo model and the solid line shows
the Monte Carlo model after applying flux reweighting. The ratio of
the data to Monte Carlo is shown below each distribution.
The error bars show the statistical and systematic uncertainties added
in quadrature.}

\label{fig:data_mc_compare}
\end{figure*}

Fig. \ref{fig:data_mc_compare} shows a comparison of the CC data sample
and Monte Carlo simulation before and after flux reweighting is applied for
the measured kinematic variables; the muon energy, hadronic shower energy, and 
muon track angle with respect to the beam direction.
The agreement between Monte Carlo and data in all three distributions
is significantly improved 
after the flux reweighting has been applied. 

\subsection{Cross Section Extraction\label{sub:Cross-Section-Extraction}}

The energy dependence of the cross section is extracted by dividing the selected CC rate \gcc{} by the measured flux \flux{} in each energy bin. 
Explicitly including the corrections described above, the cross section is
\begin{equation}
\frac{\xsec{}}{E} = \frac{1}{E} \frac{\gcc{}}{\flux} =\frac{1}{E} \left(\frac{(\ncc{}-\bcc{})/\acc{}}
{\nucor{}\times\hecor{}\times(\fsam{}-\bflux{})/\aflux{}}\right)
\label{eq:diffnux}
\end{equation}

Because our data extends above \unit[30]{GeV} we can take advantage of existing high energy precision 
measurements~\cite{ccfr_90,ccfrr,cdhsw,bebc,ggm_sps} to determine the flux normalization constant $H^{\nu}$.
\hecor{} is chosen so that our measured $\nu_\mu$ flux results in an average $\nu_\mu$ CC cross section value
from 30-\unit[50]{GeV} that agrees with the world average on an isoscalar target (0.675 $\pm$ 0.009)\unit[$\times10^{-38}$]{cm$^{2}$/GeV}
in the same energy range.  
The uncertainty on the normalization constant from the statistical precision of our data and that from the world average cross section are added 
in quadrature and applied as an uncertainty on our measured cross section at all energies. 
We apply a normalization correction to the antineutrino sample
to account for the small difference in $F_2^{\nu}$ and $F_2^{\bar{\nu}}$. The correction, which is computed from our cross section model,
is 1\% for $\nu<$\unit[1]{GeV}, 2.6\% for $\nu<$\unit[2]{GeV} and 3.8\% for the $\nu<$\unit[5]{GeV} 
sample~\cite{debdatta_thesis}.

After extracting the flux with the low-$\nu$ method, the cross section
analysis is then repeated with 
the measured flux as input.
This removes the effect of inaccuracies in the initial simulated flux on the acceptance corrections that 
are applied to both the flux and the cross section samples.
The change in cross section between the final value
and that extracted with the default simulated flux is less than 0.5\% averaged over all data points.
Since this effect is small we do not iterate the procedure further. 

The cross section we report is that expected for an isoscalar target.
The MINOS iron-scintillator detector has a 6.1\% excess of neutrons over protons for
which we correct using the {\sc neugen3} cross section model~\cite{neugen}.
The energy dependent corrections are about -2\% for neutrinos and +2\% for antineutrinos.
We also apply corrections for radiative effects~\citep{radiative}, which 
have an effect on the result of less than 1\% at all energies.

Appendix~\ref{appendix:uncorr} provides the measured raw ratio of cross 
section to flux data samples where both numerator and denominator are
corrected only for detector effects and backgrounds. As described further therein,
this allows one to remove the model dependence in the cross section extraction 
and to use an alternative cross section model to correct the ratio and 
compute cross sections.

\section{Systematic Uncertainties\label{sec:Systematics}}

Systematic uncertainties in the measurement from the following sources
were considered: reconstructed muon and hadron energy scales, the effect of final
state interactions on the measured energy, NC contamination, wrong-sign
contamination (antineutrino sample only), our lack of perfect understanding of the detector
and event reconstruction, and cross section modeling. Each systematic
uncertainty is evaluated independently and propagated through the
analysis, including a recalculation of the absolute normalization
of the result. Many systematic effects cause changes that are similar in the
cross section and flux samples and therefore partially cancel
in the measured cross section. 

The largest uncertainty comes from knowledge of the absolute muon
and hadronic energy scales discussed in Sec.~\ref{sub:Calibration}.
The muon energy scale is more important for the flux sample than for
the cross section sample because a larger fraction of the neutrino
energy per event is carried by the muon in the former. Conversely,
the hadronic energy scale is less important in the flux sample.

\begin{figure*}
\includegraphics[width=8cm]{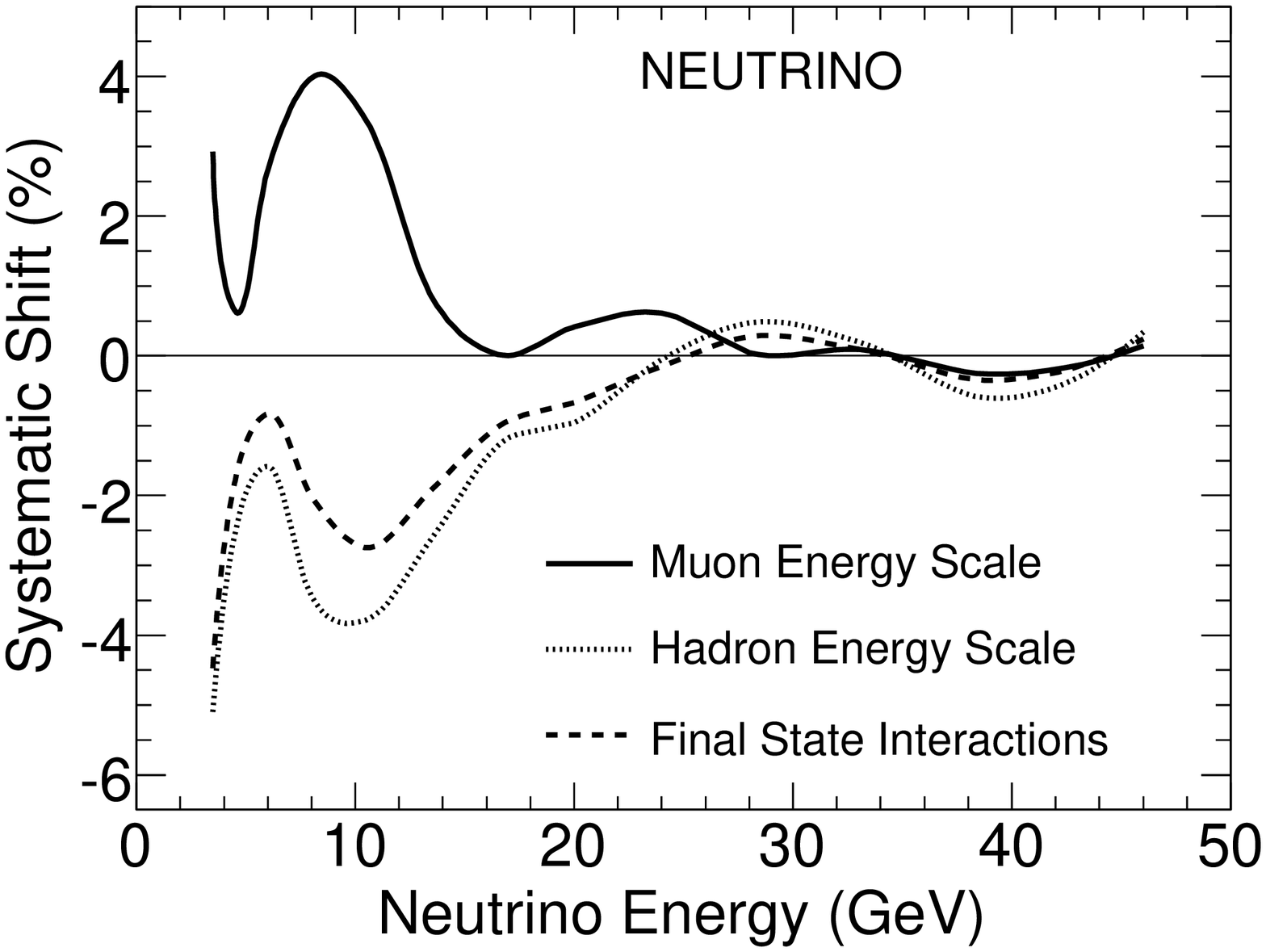}
\includegraphics[width=8cm]{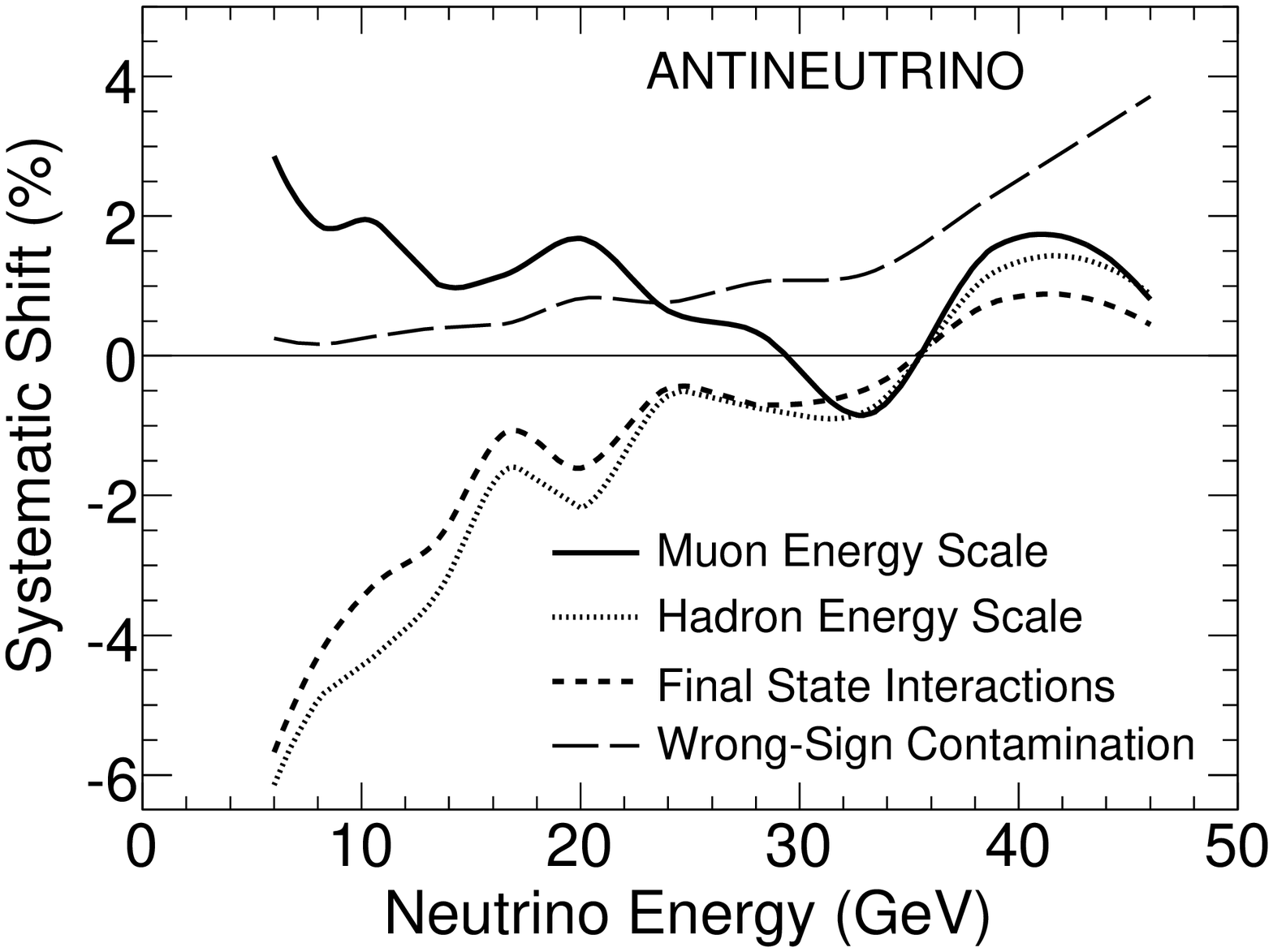}

\caption{Effect of the energy scale uncertainty on the neutrino (left) and antineutrino (right)
extracted cross section. The curves give the shape distortion due to a
one-sided one-sigma error.  The solid line shows the effect of increasing
the muon energy scale by 2\% for stopping muons and 4\% for exiting
muons. The dotted line shows the effect of increasing the hadronic
energy scale by 5.6\%, and the dashed line shows the effect of shifting the
final state interaction model. The plot on the right also shows the effect
of wrong-sign contamination uncertainty (large dashes) on the antineutrino cross section.
}

\label{fig:energy_scale}
\end{figure*}

Fig. \ref{fig:energy_scale} shows
the effects of muon and hadronic energy scale uncertainties on the
extracted cross section. These are evaluated by applying the
one-sigma shift in each scale factor to the data, extracting a new flux, 
and determining a new cross section, including a new normalization 
to the world data between 30 to \unit[50]{GeV}.  The resulting curves in 
Fig. \ref{fig:energy_scale} represent the change from the baseline cross section measurement.  
These uncertainties peak in the low energy region and become small at
high energies where the normalization is pinned.

The energy dependence of these uncertainties has a non-trivial 
shape because of the interplay of the shape of the flux spectrum and the method of 
normalizing to the world high energy measurements. 
In particular, applying a muon energy scale factor shifts the observed
flux in one direction and causes 
inflection points in the shape of the extracted flux near \unit[6]{GeV} and \unit[14]{GeV} 
for neutrinos, which propagate (with some cancellation)
to the cross section analysis as shown in 
Fig. \ref{fig:energy_scale}. 
A similar effect arises more directly in the cross section sample for the 
hadron energy scale.
We show the +1$\sigma$ systematic alone so that the shape distortion is clear.
The  -1$\sigma$ distortion is the same shape but inverted in sign.
The antineutrino analysis has fewer inflections because the spectrum
is not peaked.

As described in Sec.~\ref{sub:Simulation}, final state interactions
affect the measured hadronic energy and are included in our
Monte Carlo simulation. 
The uncertainty in their modeling contributes an effective 
hadronic energy scale uncertainty of 8\% below hadronic energy of \unit[1]{GeV}, 
decreasing to 4\% above \unit[5]{GeV}.
Their effect on the cross section, shown in Fig. \ref{fig:energy_scale},
peaks at low energy and is fractionally larger for antineutrinos,
which have a larger fraction of low hadron energy events due to 
their inelasticty distribution.

The uncertainty from our knowledge of the NC contamination is obtained
by varying the value of the minimum $E_{\mu}$ requirement, which selects
the CC sample, from its nominal value of \unit[1.5]{GeV} up to \unit[2.0]{GeV} and
down to \unit[1.0]{GeV}. The resulting change in $\sigma/E$ is small and
corresponds to a change of less than 1\%, which we take to be the NC
contamination uncertainty.

To account for uncertainties in the acceptance correction, which 
arise from modeling of detector geometry, alignment,
and magnetic field, we collected a dedicated data set with 
the detector magnetic field polarity reversed (set to focus 
positive charges) from its nominal running mode which focuses
negative charges. In this data set,  muon tracks pass through
a different region of the detector and previously focused tracks
bend away from the coil region. The flux and cross section are extracted
and are compared with their values measured in the nominal mode.
The flux sample relies more heavily on the muon track measurement 
than does the cross section sample and consequently larger effects
are seen in the measured flux (on average $5\%$ for neutrino and 2.5\% for antineutrino 
flux). Smaller differences are expected for the antineutrino sample because
more of the tracks exit the detector in the downstream region and
do not pass near the difficult-to-model coil region in either mode.
A systematic uncertainty on the nominal polarity flux is taken to be
half the difference between it and the flux extracted in the reversed polarity
sample. This is added in quadrature with the other uncertainties. 
The cross section sample relies less heavily on the track momentum.
The differences in measured cross section are at the level of 1\%, which are neglected.

Since the cross section model is used to apply a small energy dependent correction
to the flux sample (see Eq. \ref{eq:nucor}), we take into account uncertainties
in the model parameters described in Sec.~\ref{sub:Simulation}.
We account for uncertainties in the quasi-elastic and resonance contributions
to the cross section by varying the axial mass parameters, $M^{QE}_A$ and 
$M^{RES}_A$ in our model by $\pm 15\%$. The resulting effect on the 
cross section is less than $2\%$.
In this measurement, we include an additional uncertainty in the DIS 
 component of the model to account for contributions
to the $\nu$ dependence of the cross section that could affect the
flux extraction. To quantify the resulting uncertainty, we vary each parameter in
the model~\citep{bodek-yang} and study the change of the reduced
$\chi^{2}$ of the fit to the charged lepton data from which they were
originally determined. We take the shift that corresponds to a one
unit shift in fit $\chi^{2}$ as the uncertainty for each parameter.
The values and the associated uncertainties of $A_{ht}$, $B_{ht}$,
$C_{v1u}$ and $C_{v2u}$ are determined to be 0.538$\pm$0.134, 0.305$\pm$0.076,
0.291$\pm$0.087 and 0.189$\pm$0.076, respectively. The other parameters ($C_{v1d}$,
$C_{v2d}$, $C_{su}$ and $C_{sd}$) have a negligible effect on
the analysis.  The effects of these uncertainties on the flux measurement
and the acceptance correction
are propagated to an uncertainty on the extracted cross section. The contributions 
from each parameter shift are added linearly to form the total DIS model uncertainty
which is  2\% below \unit[8]{GeV} for both neutrino and antineutrino
cross sections, and is negligible above this energy. 

The contamination from wrong-sign events is significant only for
the antineutrino sample. To evaluate the uncertainty from this source,
we recompute the cross section assuming no wrong-sign contamination and two times
as much wrong-sign contamination. The resulting
uncertainty is negligible below \unit[15]{GeV} but is about 4\%
at the highest energies. 

\begin{figure*}
\includegraphics[width=8cm]{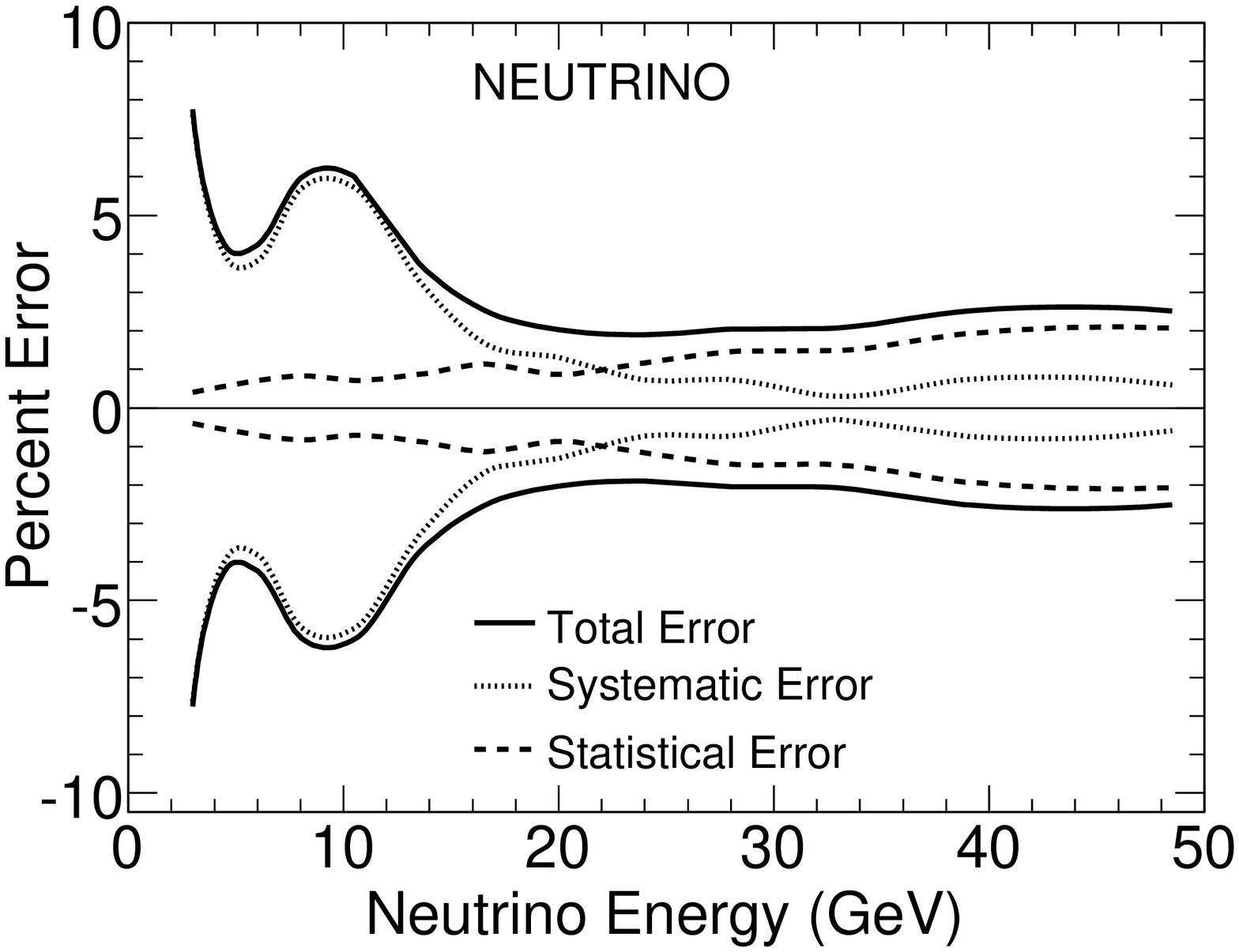}
\includegraphics[width=8cm]{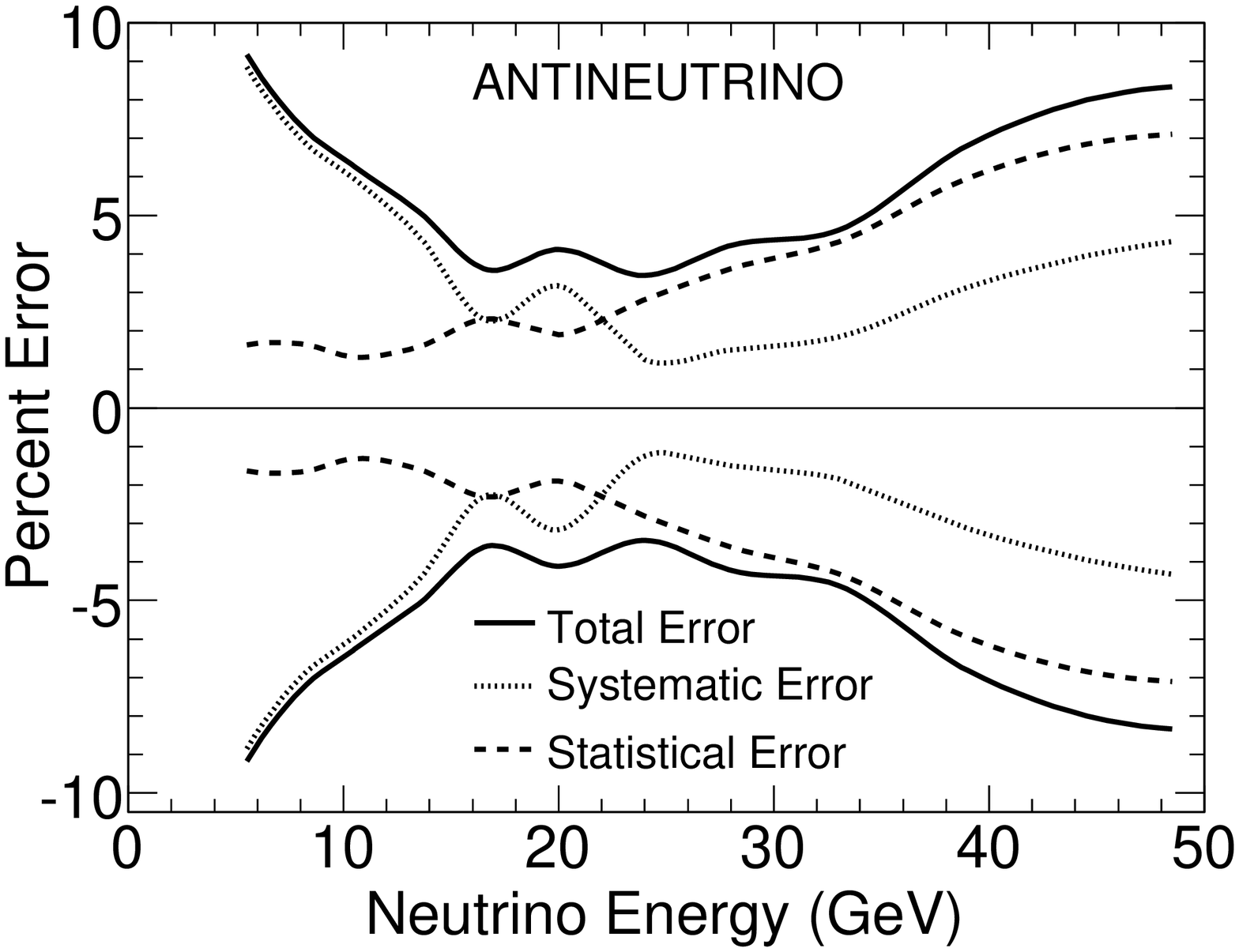}

\caption{Summary of the statistical, total systematic, and total uncertainty
for the neutrino (left) and antineutrino (right) extracted cross section.
}

\label{fig:total_error}
\end{figure*}

The systematic, statistical, and total uncertainties, for the neutrino and antineutrino
cross sections are summarized in Fig. \ref{fig:total_error}.

In the cross section ratio
significant additional cancellation of uncertainties occurs.  At the 
lower energies the uncertainties are about half those shown in 
Fig. \ref{fig:energy_scale}.
At higher energies they are one to two percent each,
except for the wrong-sign contamination which is significant
only for the antineutrino sample. 

%

\section{Results}
\label{sec:Results}

\begin{figure*}
\includegraphics[width=0.5\textwidth]{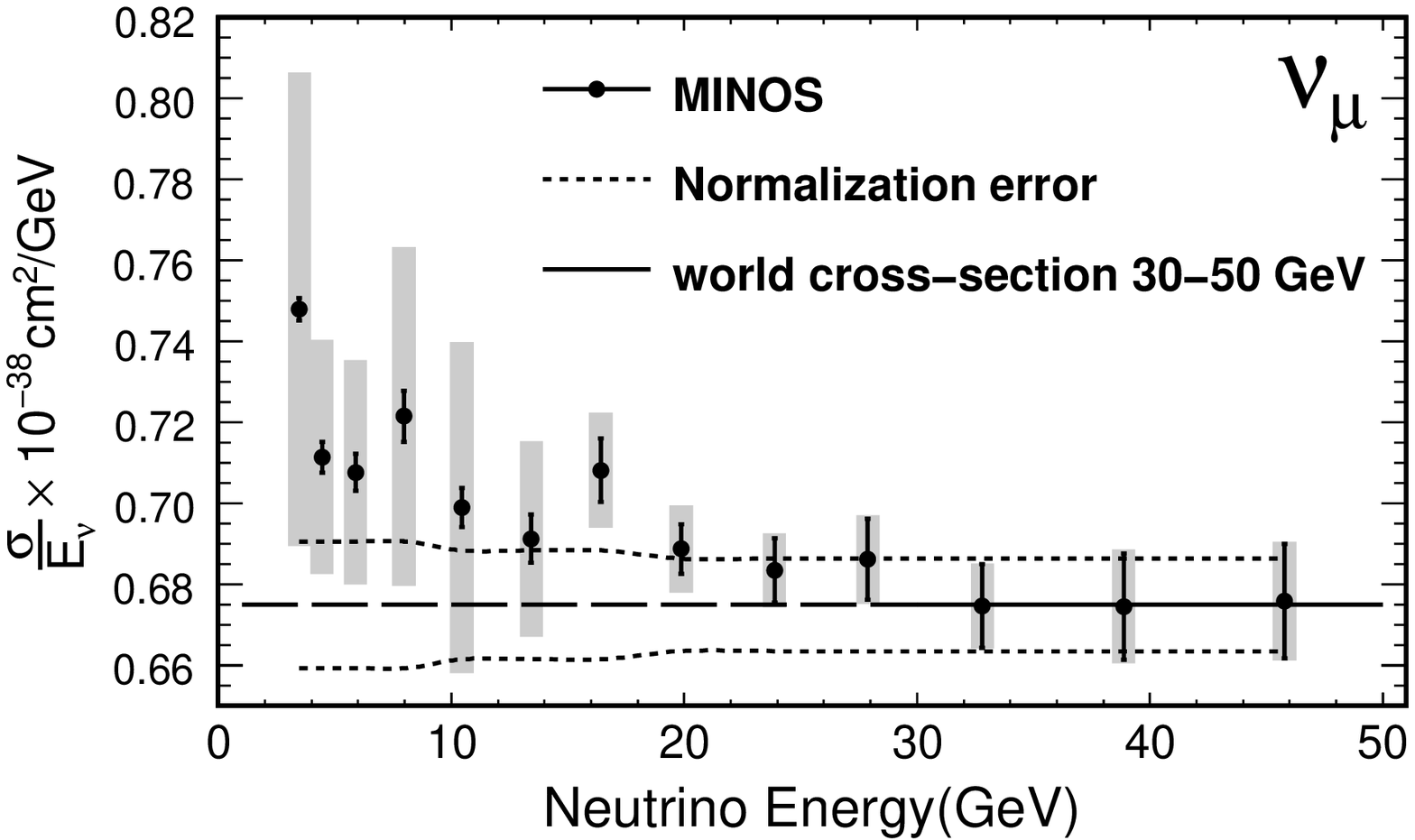}\includegraphics[width=0.5\textwidth]{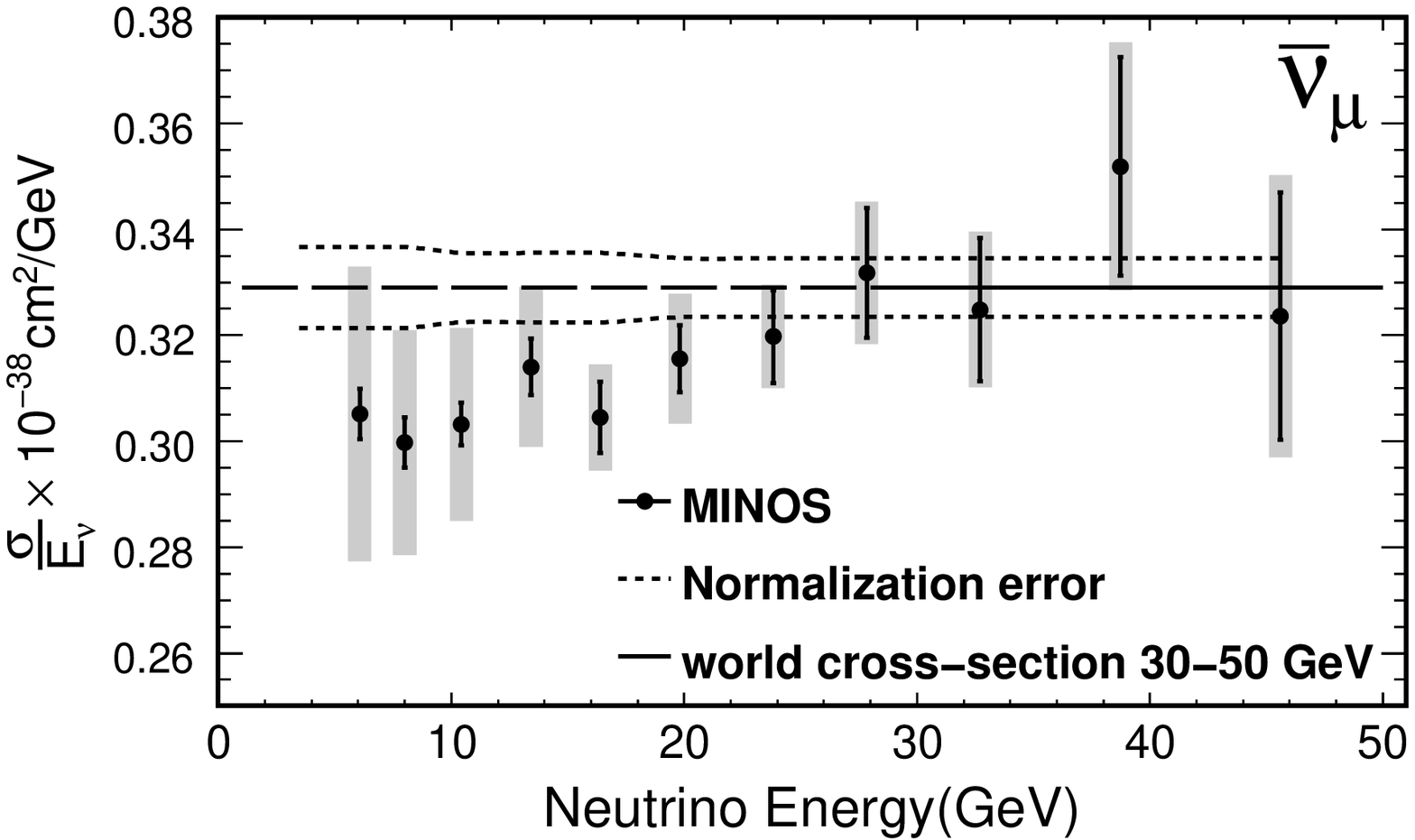}

\caption{Neutrino (left) and antineutrino (right) charged-current inclusive cross section per nucleon divided by energy for an isoscalar iron target. 
The black error bars show the statistical uncertainty and the shaded boxes show the statistical and systematic uncertainties added in quadrature. The 
dotted band shows the uncertainty on the normalization (about 1.5\%). The solid black line shows the world average neutrino cross section value of 
\unit[$0.675\times10^{-38}$]{cm$^{2}$/GeV} from 30 to \unit[50]{GeV}~\cite{ccfr_90,ccfrr,cdhsw,bebc,ggm_sps} and the dashed black line shows this 
value extrapolated to lower energies. The neutrino cross section above \unit[30]{GeV} is normalized using this world average value and the same 
normalization constant is then applied to the antineutrinos. The solid black line on the antineutrino cross section plot shows a world average 
antineutrino cross section value of \unit[$0.329\times10^{-38}$]{cm$^{2}$/GeV} from 30 to \unit[50]{GeV}~\cite{ccfrr,cdhsw,ggm_sps}. This value is shown for 
comparison and is not used for antineutrino sample normalization.}

\label{fig:nu_xsec}
\end{figure*}

Fig. \ref{fig:nu_xsec} shows the extracted energy dependence of
the total cross section divided by energy ($\sigma/E$)
for $\nu_{\mu}N$ CC and for $\bar{\nu}_{\mu}N$ CC 
interactions on an isoscalar target.
The cross section values are assigned to the average energy in the bin.
Both cross sections approach a linear energy dependence for energies above \unit[20]{GeV}.
For neutrinos $\sigma/E$ drops with increasing energy 
in the lower energy region.
At \unit[3]{GeV} the quasi-elastic cross section is still expected to be appreciable 
($\approx$15\%). Its contribution to $\sigma/E$
 falls rapidly with increasing energy
as inelastic processes (resonance production 
and DIS) turn on.
For antineutrinos the measured $\sigma/E$ rises gradually in the region 5-\unit[20]{GeV}
to its asymptotic high energy value. In this case the falling fractional contribution of the 
quasi-elastic cross section is offset by the more gradual turn on of the 
DIS process, which is expected due to its strong dependence on the antiquark component
which rises slowly with increasing $Q^2$. 
Table~\ref{tab:nu_xsec_table} summarizes the neutrino and 
antineutrino cross section results.

\begin{table*}
\begin{tabular}{|c||c|c|c|c|c|c||c|c|c|c|c|c|}
\hline 
  &  \multicolumn{6}{c||}{Neutrino} &  \multicolumn{6}{c|}{Antineutrino} \\
\hline 
$E$ bin & $<E_\nu>$ & $\sigma$/$E$ & stat. & sys. & norm. & total &
$<E_{\overline{\nu}}>$ & $\sigma$/$E$ & stat. & sys.  & norm. & total  \\
 & &  & error & error & error & error  &  & & error & error & error & error \\
\hline
\hline 
\multicolumn{2}{|c|}{(GeV)} & \multicolumn{5}{c||}{($10^{-38}$cm$^{2}$/GeV)} &
(GeV) & \multicolumn{5}{c|}{($10^{-38}$cm$^{2}$/GeV)}\tabularnewline
\hline 
3-4 & 3.48  & 0.748  & 0.003  & 0.058  & 0.017  & 0.061 & \multicolumn{6}{c|}{ }\tabularnewline
\hline 
4-5 & 4.45  & 0.711  & 0.004 & 0.029 & 0.017  & 0.033 & \multicolumn{6}{c|}{ }\tabularnewline
\hline 
5-7 & 5.89  & 0.708  & 0.005  & 0.027 & 0.016  & 0.032 &
6.07  & 0.305  & 0.005  & 0.027  & 0.007 &  0.029 \tabularnewline
\hline 
7-9 & 7.97  & 0.722 & 0.006  & 0.041  & 0.017  & 0.045 &
7.99  & 0.300  & 0.005  & 0.021  & 0.007  & 0.022 \tabularnewline
\hline 
9-12 & 10.45  & 0.699  & 0.005  & 0.041  & 0.014  & 0.043 &
10.43  & 0.303  & 0.004  & 0.018  & 0.006  & 0.019 \tabularnewline
\hline 
12-15 & 13.43  & 0.691  & 0.006  & 0.023  & 0.014  & 0.028 &
13.42  & 0.314  & 0.005  & 0.014  & 0.006  & 0.016 \tabularnewline
\hline 
15-18 & 16.42  & 0.708  & 0.008  & 0.012  & 0.014  & 0.020 &
16.41  & 0.304  & 0.007  & 0.007  & 0.006  & 0.012\tabularnewline
\hline 
18-22 & 19.87  & 0.689  & 0.006  & 0.009  & 0.012  & 0.016 &
19.82  & 0.316  & 0.006  & 0.011  & 0.005  & 0.013\tabularnewline
\hline 
22-26 & 23.88  & 0.683  & 0.008  & 0.005  & 0.012  & 0.015 &
23.82  & 0.320  & 0.009  & 0.004  & 0.005  & 0.011\tabularnewline
\hline 
26-30 & 27.89  & 0.686  & 0.010  & 0.004  & 0.012  & 0.016 &
27.84  & 0.332  & 0.012  & 0.005  & 0.006  & 0.015\tabularnewline
\hline 
30-36 & 32.81  & 0.675  & 0.010  & 0.002  & 0.011  & 0.016 &
32.72  & 0.325  & 0.014  & 0.006  & 0.005  & 0.016\tabularnewline
\hline 
36-42 & 38.87  & 0.675  & 0.013  & 0.005  & 0.011  & 0.018 &
38.74  & 0.352  & 0.021  & 0.011  & 0.006  & 0.024\tabularnewline
\hline 
42-50 & 45.77  & 0.676  & 0.014  & 0.004  & 0.011  & 0.019 &
45.61  & 0.324  & 0.023  & 0.013  & 0.005  & 0.027\tabularnewline
\hline
\end{tabular}
\caption{Summary of neutrino (left) and antineutrino (right) cross section results. 
The second column for each species shows the average energy in each bin. 
The uncertainties shown in columns 4-7 for each species are the statistical, 
systematic, normalization and total contributions, respectively.
The total uncertainty is obtained by summing the statistical, systematic, and normalization 
uncertainties in quadrature. }
\label{tab:nu_xsec_table}
\end{table*}

\begin{figure*}
\includegraphics[width=16cm]{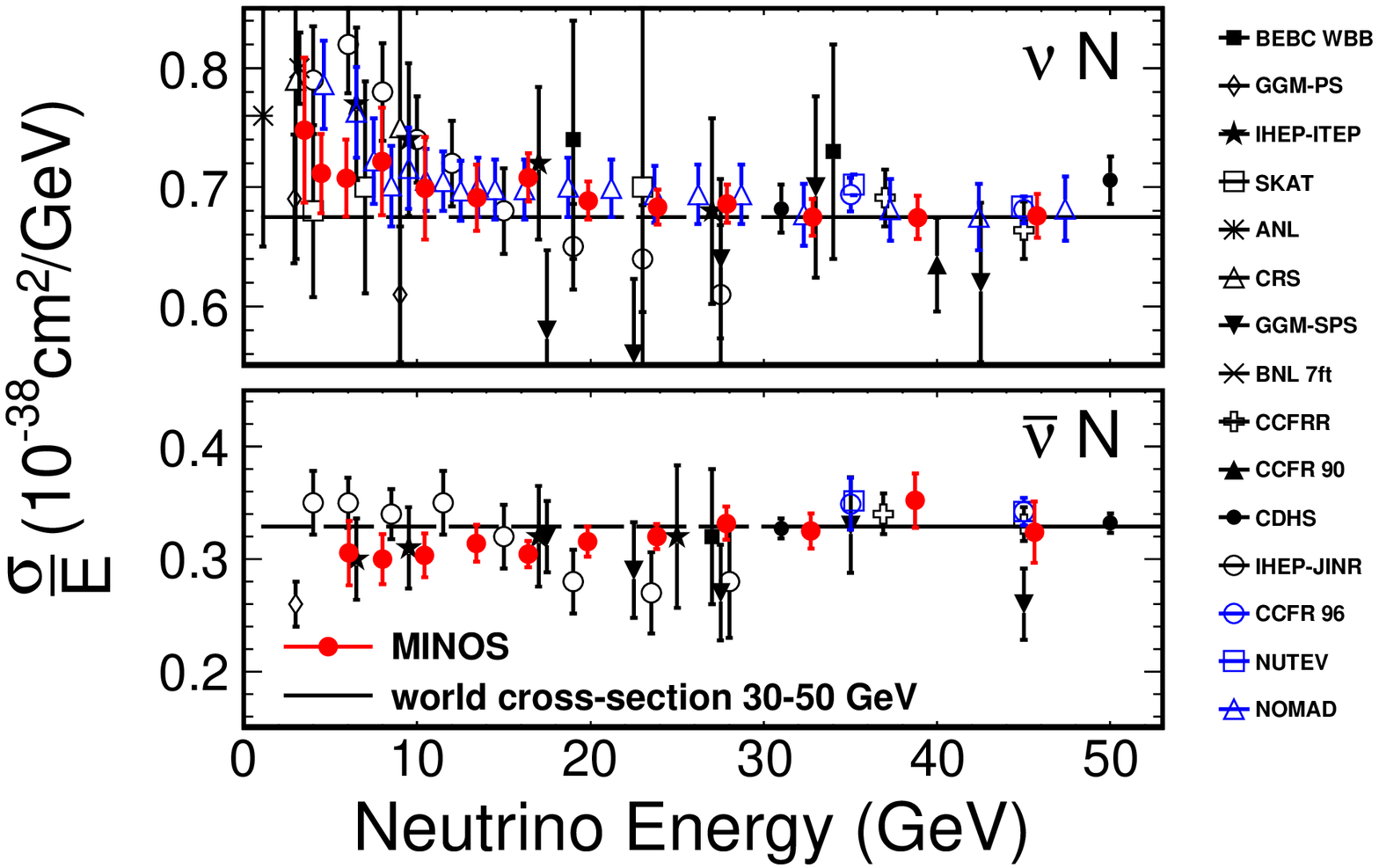}

\caption{MINOS neutrino and antineutrino charged-current inclusive cross section
compared with other experimental 
results~\citep{anl,bebc,bnl,ccfr_90,ccfrr,cdhsw,charm,crs,ggm-ps-anu,ggm_ps,ggm_sps,ihep_itep,ihep_jinr,nomad,nutev,seligman}. 
The error bars show the statistical, systematic, and normalization
uncertainties added in quadrature. The solid black line shows the
average world cross section in the 30 to \unit[50]{GeV} region for the neutrino 
(\unit[$0.675\times 10^{-38}$]{cm$^{2}$/GeV}) and the antineutrino (\unit[$0.329 \times 10^{-38}$]{cm$^{2}$/GeV}). 
The dashed line shows these high energy values extrapolated to lower energies.}

\label{fig:world_data}
\end{figure*}

Fig. \ref{fig:world_data} shows MINOS neutrino and antineutrino
results compared to the results from other experiments. The MINOS
neutrino cross section agrees with previous measurements from CRS~\citep{crs},
SKAT~\citep{skat}, IHEP-JINR~\citep{ihep_itep}, and GGM-PS~\citep{ggm_ps}, but these experiments
have significantly larger uncertainties. 
Our neutrino cross section 
is in good agreement and has comparable precision with the recent NOMAD measurement~\cite{nomad}.
Our result is systematics-limited in the region below \unit[15]{GeV} where the
largest uncertainties comes from knowledge of the absolute muon and hadronic energy scales, whereas
in NOMAD the flux determination dominates the uncertainty. 
The MINOS antineutrino cross section result is in good agreement with the
sparse data available at lower energies and has much smaller uncertainty
in the 10 to \unit[30]{GeV} region.

\begin{figure*}
\includegraphics[width=0.5\textwidth]{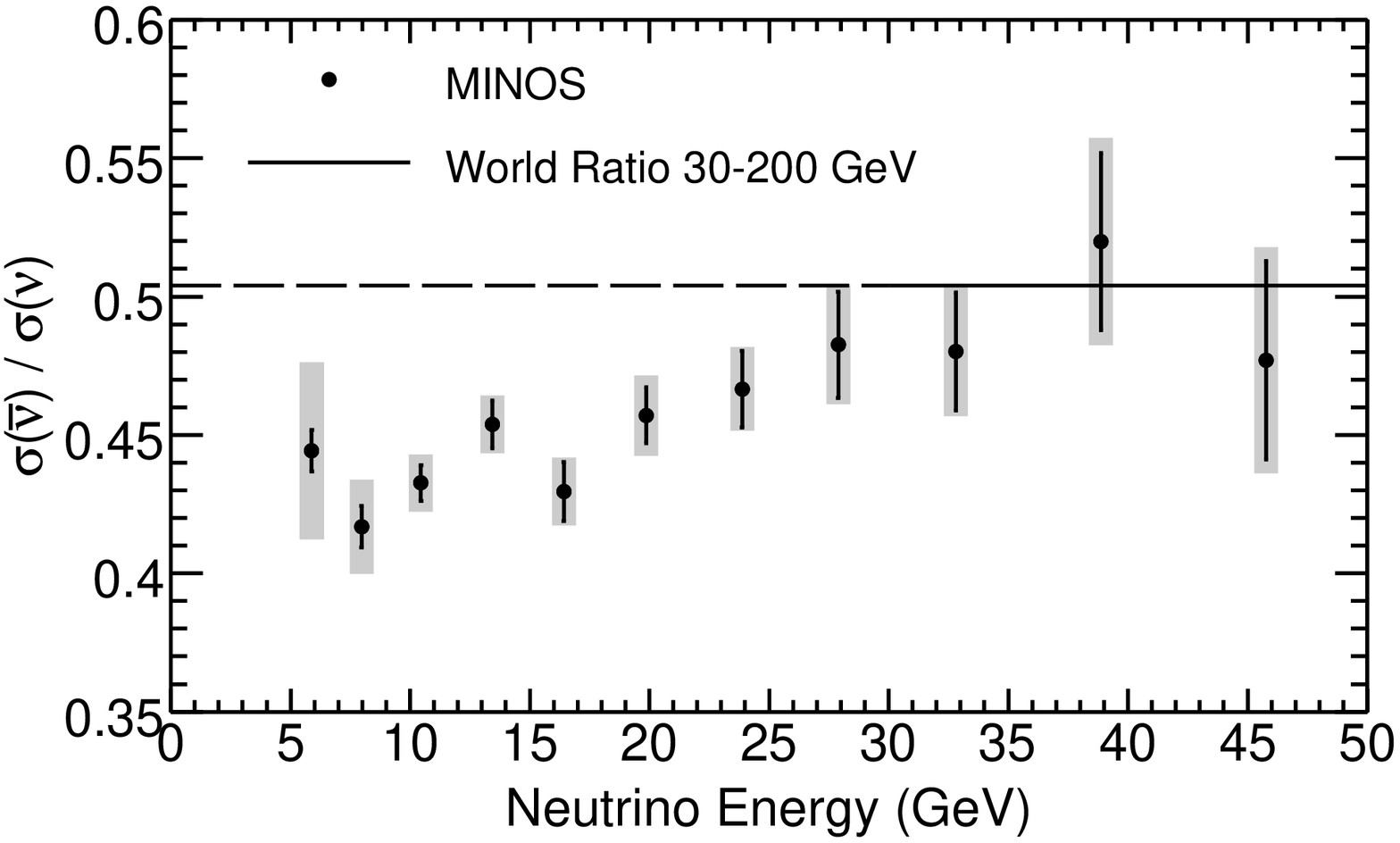}\includegraphics[width=0.5\textwidth]{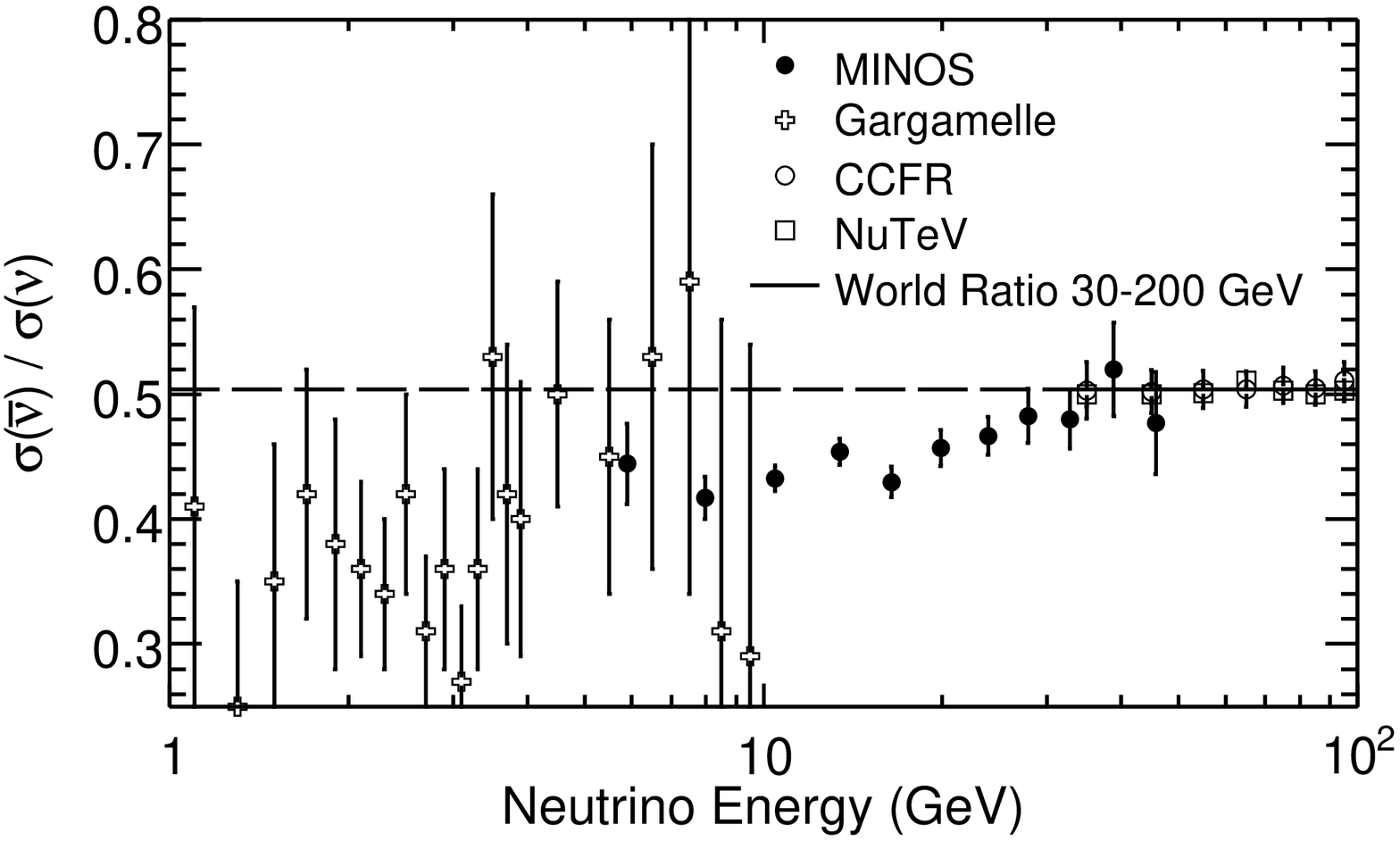}
\caption{(left) Ratio of antineutrino-nucleon to neutrino-nucleon cross section as
a function of energy. Black error bars show the statistical uncertainty
and shaded boxes show the total uncertainty with statistical and systematic uncertainties
added in quadrature. The solid black line at 0.504$\pm$0.003 is drawn at the 
average value obtained from previous measurements over the energy range 30-\unit[200]{GeV}~\cite{rpp}.
(right) Comparison of measured $r$ with other measurements 
for $E<$\unit[100]{GeV}. The MINOS result 
spans the intermediate energy range and overlaps  
with the low energy data~\citep{eichten} as well as with 
precise high energy measurements~\citep{nutev,seligman}. }

\label{fig:littler}
\end{figure*}

Fig. \ref{fig:littler} and Table \ref{tab:r_table}
show the ratio of the $\bar{\nu}_{\mu}$N CC to $\nu_{\mu}$N CC inclusive cross section as a function of energy.
Because of cancellation of many of the systematic uncertainties the  MINOS
result is statistics-limited above \unit[10]{GeV}.
The cross section ratio appears to gradually
approach its asymptotic scaling value of  0.504$\pm$0.003, 
defined by the world average from 30-\unit[200]{GeV} calculated from 
previous experiments~\cite{rpp}. 
The MINOS average ratio measured from 30-\unit[50]{GeV} of 0.489$\pm$0.012 
is in good agreement with the asymptotic value. 
At \unit[10]{GeV} the measured ratio is 14\% below the asymptotic
value with 6.6$\sigma$ significance and at \unit[24]{GeV} the measurement lies 7\% below
with 2.4$\sigma$ significance. Our precise data show 
a slower approach to scaling behavior than has been previously  claimed by
low energy measurements\footnote{These are based on fits to the measured
neutrino cross section with energy.},
which found their data to be consistent with scaling in the few GeV range~\citep{bnl,skat,eichten}.

Fig. \ref{fig:littler} also shows the cross section ratio compared with 
the few other existing measurements.
The MINOS data uniquely spans the 10-\unit[30]{GeV} region.  
It overlaps the precise high energy measurements~\citep{nutev,seligman}
as well as the  
Gargamelle low energy measurement~\citep{eichten}
which has precision of only about 20\%.

The total neutrino and antineutrino cross section in the Quark
Parton Model, which describes neutrino scattering at high energy, 
can be written as
\begin{equation}
\sigma(\nu N)=\frac{G_{F}^{2}ME}{\pi}(Q+\frac{1}{3}\bar{Q}),\label{eq:result1}\end{equation}
and
\begin{equation}
\sigma(\overline{\nu}N)=\frac{G_{F}^{2}ME}{\pi}(\overline{Q}+\frac{1}{3}Q)\label{eq:result2}
\end{equation}
with the ratio of the two given by
\begin{equation}
r=\frac{\sigma(\overline{\nu} N)}{\sigma(\nu N)}=\frac{1+3\bar{Q}/Q}
{3+ \bar{Q}/Q}, \label{eq:qpmr}
\end{equation}
where $Q=\int x[u(x)+d(x)]dx$ 
and $\overline{Q}=\int x[\overline{u}(x)+\overline{d}(x)]dx$.
Here $u(x)$ ($\overline{u}(x)$) and  $d(x)$ ($\overline{d}(x)$) are the parton distribution 
functions (PDFs) for the up and down quarks (antiquarks) in the nucleon, respectively.
In the limit of large  $Q^{2}$, $Q^{2}>>M^2$,
the PDFs depend only on $x$ and are independent of $Q^{2}$. 
In this limit the QPM predicts 
scaling behavior, {\em i.e.}, a linear dependence 
of the cross sections with energy. 
In the low energy (low  $Q^{2}$) limit,
scaling violations occur and the QPM breaks down.
Scattering off the entire nucleon
(quasi-elastic scattering) and resonance production, where the nucleon 
is excited and decays to low multiplicity final states, must also be
considered to account for the energy dependence of the cross section.

The ratio $r$ is constant with energy in the QPM
and depends only on the integrated quark and antiquark distributions in the 
high-$Q^{2}$ limit. 
Eq. \ref{eq:result2} indicates antiquarks are relatively more important in the 
antineutrino scattering case.
$r$ approaches the limiting value of 1/3 if antiquarks are not
present in the nucleon. High energy measurements of $r$ can be used to measure 
the fraction of momentum carried by antiquarks in the nucleon. 

In order to interpret our measurement of $r$ in the context of the QPM 
the quasi-elastic contribution is removed from the measured value
by defining  $r_{inel}=\sigma^{\overline{\nu}}_{inel}/\sigma^{\nu}_{inel}$ to be
the cross section ratio for the purely inelastic contribution to the cross section.
To compute $r_{inel}$, the {\sc neugen3} cross  section model~\citep{neugen} is
used to remove the fractional quasi-elastic contribution\footnote {The 
{\sc neugen3} model, as described earlier, uses a value of 
$M^{QE}_A=0.99$. 
We provide the raw measured $r$ so that one can use
other models to compute the inelastic fraction. This will be especially useful as
knowledge of $M^{QE}_A$ improves. For reference, increasing  $M^{QE}_A$ by $0.15$ decreases
the inelastic fraction by less than 1\% at \unit[5.9]{GeV},
which is small compared with the experimental uncertainty.}.
%
Table \ref{tab:r_table} gives the measured ratio $r$ and the inelastic fraction  $r_{inel}$
along with their experimental uncertainties. The similarly slow increase of
$r_{inel}$ with energy shows that the decrease in the quasi-elastic
contributions alone has only a small effect on the observed shape.

\begin{table*}
\begin{tabular}{|c|c|c|c||c|c|c|c|}
\hline 
Energy (GeV) & r & Stat. err. & Total err. & $r_{inel}$ & $\Delta r_{inel}$ & $\frac{\bar{Q}}{Q+\bar{Q}}$ &  $\Delta (\frac{\bar{Q}}{Q+\bar{Q}})$\\ \hline \hline
5.9 & 0.444 & 0.007 & 0.032  & 0.407 & 0.029 & 0.079 & 0.030 \\ \hline
8.0 & 0.417 & 0.008 & 0.016  & 0.389 & 0.016 & 0.060 & 0.016 \\ \hline
10.5 & 0.433 & 0.006 & 0.010 & 0.410 & 0.010 & 0.081 & 0.010  \\ \hline
13.4 & 0.454 & 0.009 & 0.010 & 0.435 & 0.010 & 0.106 & 0.010  \\ \hline
16.4 & 0.430 & 0.011 & 0.012 & 0.415 & 0.012 & 0.086 & 0.012  \\ \hline
19.9 & 0.457 & 0.010 & 0.015 & 0.444 & 0.014 & 0.115 & 0.014  \\ \hline
23.9 & 0.467 & 0.014 & 0.015 & 0.455 & 0.015 & 0.126 & 0.014 \\ \hline
27.9 & 0.482 & 0.019 & 0.022 & 0.472 & 0.021 & 0.142 & 0.019  \\ \hline
32.8 & 0.480 & 0.021 & 0.023 & 0.472 & 0.023 & 0.141 & 0.021  \\ \hline
38.9 & 0.520 & 0.032 & 0.037 & 0.512 & 0.037 & 0.177 & 0.032  \\ \hline
45.8 & 0.477 & 0.036 & 0.041 & 0.471 & 0.040 & 0.140 & 0.037 \\ \hline
\end{tabular}
\caption{The measured cross section ratio $r$ at the bin average energy along with statistical and 
total uncertainties are given.  To compute $r_{inel}$, the  {\sc neugen3}~\citep{neugen} cross section 
model is used to remove the fractional quasi-elastic contribution. The quark parton model is used
to estimate the fraction of the total quark momentum that is carried by antiquarks, $\frac{\bar{Q}}{Q+\bar{Q}}$.
Uncertainties computed for $\Delta r_{inel}$ and  the antiquark fraction
do not include any model uncertainty contributions.}
\label{tab:r_table}
\end{table*}

Eq. \ref{eq:qpmr} can be rearranged to give the fraction of total quark momentum in the
nucleon that is carried by antiquarks,
$\frac{\bar{Q}}{Q+\bar{Q}}=\frac{1}{2}\frac{(3r-1)}{(r+1)}$.
This fraction as a function of energy is also given in Table \ref{tab:r_table}.
As neutrino energy decreases, 
one moves increasingly away from the domain of validity of this expression,
which is derived in the DIS region.
Target mass corrections as well as higher-twist terms become more
important, especially at high-$x$. However, the high-$x$ region contributes little
to the $Q$ and $\bar{Q}$ integrals.
In the approximation that 
the contributions from these effects are small, 
our results are consistent with a non-zero antiquark content in the 
nucleon at our lowest energy, \unit[5.9]{GeV} (\unit[$\langle Q^2 \rangle = 1.4$]{GeV$^2$}) and a gradual increase of
the antiquark fraction with energy.
In order to accurately extract the antiquark
fraction from our data, a full higher order QCD model that
incorporates these effects is required.

\section{Conclusion}

We have measured the charged-current neutrino-nucleus inclusive cross section
in the energy range 3-\unit[50]{GeV} with a precision of 2-8\% and
the antineutrino-nucleus cross section from 5-\unit[30]{GeV} with a precision in the
range 3-9\%. The flux was determined by using 
a subsample of low-hadronic-energy events to measure the flux shape and 
the world average cross section above \unit[30]{GeV} for normalization.
This method was previously used at higher energies~\citep{nutev,seligman}
and here we have extended it down to \unit[3]{GeV}. 
While the measurements are systematics-dominated, the overall
systematic uncertainty benefits from partial cancellation in detector
related systematic uncertainties that arise from measuring the 
flux and the CC event rate in the same detector.
Both measurements impact the precision of total cross section measurements
in the less than \unit[30]{GeV} range. 

Our measurement of the antineutrino to neutrino 
cross section ratio is the most precise in the less than \unit[30]{GeV} range, where only
one previous measurement has been performed~\citep{eichten}. 
The measured rise of the cross section ratio with energy is consistent 
with an expected slow rise in the antineutrino inelastic cross section 
with the increase in number of sea-quark degrees of freedom for increasing
$Q^2$.

The measurement presented here can be used to tune neutrino and antineutrino cross section models 
which benefit ongoing and future neutrino oscillation measurements. 

\begin{acknowledgments}
This work was supported by the US DOE; the UK STFC; the US NSF; the
State and University of Minnesota; the University of Athens, Greece;
and Brazil's FAPESP and CNPq.  We are grateful to the Minnesota
Department of Natural Resources, the crew of the Soudan Underground
Laboratory, and the staff of Fermilab for their contribution to this
effort.
\end{acknowledgments}

\appendix
\noindent
\section*{Appendix: Uncorrected Data Sample}
\label{appendix:uncorr}

Our measurement has cross section model dependence which arises from the
correction for the minimum muon energy requirement $E_\mu>$\unit[1.5]{GeV} in the 
cross section sample (\ncc{}) and for
the small energy dependence in the low-$\nu$ flux sample (see Eq.~\ref{eq:dsigmadnu}). 
Here we provide the raw data cross section to flux sample ratio $R^{{\nu}(\overline{\nu})}$
corrected only for detector effects and backgrounds.
This will allow the reader to use an alternative 
cross section model and our data 
to compute neutrino and antineutrino cross sections.

Table \ref{table:xsec_raw_ratio} gives the values of $R^{\nu}$
($R^{\overline{\nu}}$ for antineutrinos) where both numerator and 
denominator have been corrected for detector effects and backgrounds using our {\sc geant3}-based
detector simulation. The ratio has not been corrected for the effect
of the kinematic $E_\mu>$\unit[1.5]{GeV} cut, which affects only the numerator.
The unnormalized neutrino cross section $\sigma^{\nu}_{unnorm}(E)$
can be computed from $R^{\nu}(E)$ 
by applying two corrections,
\begin{equation}
\sigma^{\nu}_{unnorm}(E)= K^{\nu}(E)\times 
S^{\nu}(\nu_0,E)\times R^{\nu}(E), 
\label{eq:newxsec}
\end{equation}
where $K^{\nu}(E)=N(E)/N(E,E_{\mu}>$\unit[1.5]{GeV}) is the ratio of total cross section events for all 
muon energies to that with muon energies larger than \unit[1.5]{GeV} in each energy bin, 
and $S^{\nu}(\nu_0,E)$
is defined in Eq.~\ref{eq:nucor}, (where $\nu_0$=1,2 or \unit[5]{GeV}).
\begin{table*}
\begin{tabular}{|c||c|c||c|c|}
\hline 
  &  \multicolumn{2}{c||}{Neutrino} &  \multicolumn{2}{c|}{Antineutrino} \\
\hline 
$E$ bin~(GeV) & $\langle E_\nu \rangle$~(GeV) & ~~~$R^{\nu}$~~~  & 
$\langle E_{\overline{\nu}} \rangle$~(GeV) & ~~~$R^{\overline{\nu}}$~~~ \\
\hline
\hline 
3-4 & 3.48  & 1.72  &  & \tabularnewline
\hline 
4-5 & 4.45  & 2.35  & & \tabularnewline
\hline 
5-7 & 5.89  & 3.31 & 6.07  & 2.22  \tabularnewline
\hline 
7-9 & 7.97  & 4.93 & 7.99  &  2.78 \tabularnewline
\hline 
9-12 & 10.5  & 3.71 & 10.4  & 2.10  \tabularnewline
\hline 
12-15 & 13.4  & 4.87  & 13.4  &  2.69  \tabularnewline
\hline 
15-18 & 16.4  &6.22  & 16.4  & 3.11 \tabularnewline
\hline 
18-22 & 19.9  & 3.32  & 19.8  & 1.90  \tabularnewline
\hline 
22-26 & 23.9  &3.98   & 23.8  & 2.22 \tabularnewline
\hline 
26-30 & 27.9  & 4.68 & 27.8  &  2.60  \tabularnewline
\hline 
30-36 & 32.8  &5.45  & 32.7  & 2.94 \tabularnewline
\hline 
36-42 & 38.9  & 6.49 & 38.7  & 3.75 \tabularnewline
\hline 
42-50 & 45.8  & 7.70 & 45.6  & 4.04  \tabularnewline
\hline
\end{tabular}
\caption{Ratio of cross section to flux sample where both numerator and 
denominator have been corrected for detector effects and backgrounds.
The text describes how to use this data and a cross section model to compute  
neutrino and antineutrino cross sections.}
\label{table:xsec_raw_ratio}
\end{table*}


The cross section is normalized using the values of $R^{\nu}$
for different $\nu_0$ cut samples provided in Table~\ref{table:do_norm}.
To improve statistical precision of the flux sample three different values of the 
$\nu_0$ cut were used in the analysis; $\nu_0<$\unit[1]{GeV} applies for the 
neutrino energies $E<$\unit[9]{GeV}, $\nu_0<$\unit[2]{GeV} 
for 9$<E<$\unit[18]{GeV}, and $\nu_0<$\unit[5]{GeV} for $E>$\unit[18]{GeV}.
We define the normalization constant
$Norm (\nu_0)$ for each $\nu_0$ sample as
\begin{equation}
\sigma^{\nu}_{norm}(E)=Norm (\nu_0) \times \sigma^{\nu}_{unnorm}(E).
\label{eq:newnorm}
\end{equation}
$Norm (\nu_0)$ is obtained by first computing unnormalized cross sections in the range
30$<E_\nu<$\unit[50]{GeV} using $R^{\nu}$ from Table~\ref{table:do_norm} and 
Eq.~\ref{eq:newxsec}.
The weighted average of $\sigma^{\nu}_{unnorm}(E)/\langle E_\nu\rangle$ in this energy
range is computed for each $\nu_0$ sample using the statistical errors on 
$R^{\nu}$ also given in Table~\ref{table:do_norm}.
$Norm (\nu_0)$ is then obtained by scaling this to the
world average value 0.675\unit[$\times10^{-38}$]{cm$^{2}$/GeV}.

%

\begin{table*}
\begin{tabular}{|c|c||c|c|c|}
\hline 
 $E_\nu$ (GeV) &  $<E_\nu>$ (GeV) &  \multicolumn{3}{|c|}{$R^{\nu}$} \\ \hline
   &  &  ~~~$\nu_0<$\unit[1]{GeV}~~~  &  ~~~$\nu_0<$\unit[2]{GeV}~~~ &  ~~~$\nu_0<$\unit[5]{GeV}~~~ \\ \hline
30-36 & 32.8 & 22.1 $\pm$ 0.6 & 12.3 $\pm$ 0.3 & 5.5 $\pm$ 0.1\\ \hline
36-42 & 38.9 & 26.0 $\pm$ 0.9 & 14.5 $\pm$ 0.4 & 6.5 $\pm$ 0.1\\ \hline
42-50 & 45.8 & 30.5 $\pm$ 1.2 & 17.2 $\pm$ 0.5 & 7.7 $\pm$ 0.2 \\ \hline
\hline
\end{tabular}
\caption{Raw ratio of cross section to flux sample 
(as in Table~\ref{table:xsec_raw_ratio})
in the normalization region  (30$<$E$<$\unit[50]{GeV}).
The separate $R^\nu$ columns give the ratio for the three different $\nu_0$ cut values.
Each sample is separately normalized
using the corresponding E$>$\unit[30]{GeV} data points.}
\label{table:do_norm}
\end{table*}

The same overall normalization constants are used to obtain the antineutrino
cross section 
\begin{equation}
\sigma^{\overline{\nu}}_{norm}(E)= \left[Norm(\nu_0) \times G^{corr}(\nu_0) \right] 
\times K^{\overline{\nu}}(E)\times S^{\overline{\nu}}(\nu_0,E)\times R^{\overline{\nu}}(E).
\label{eq:newxsec_nub}
\end{equation}
The additional normalization factor $G^{corr}(\nu_0)$, is used to account for 
a small difference in neutrino and antineutrino $F_2$ structure functions
(see Sec.~\ref{sub:Flux-Extraction}). $G^{corr}(\nu_0)$ can be computed 
from the ratio of asymptotic values of antineutrino to neutrino low-$\nu$ cross sections,
\begin{equation}
\label{eq:f2corr}
G^{corr}(\nu_0)=
\frac{\sigma^{\overline{\nu}}(\nu<\nu_0,E\rightarrow\infty)}{\sigma^{\nu}(\nu<\nu_0,E\rightarrow\infty)}
\end{equation}
for the three different $\nu_0$ values.

The uncertainty in the new measured cross section should be estimated using the fractional 
cross section uncertainty (syst$\oplus$stat) given in Table~\ref{tab:nu_xsec_table}, which 
properly takes into account cancellations in several systematic uncertainties in the cross 
section and flux samples.


\bibliography{bibliography}

\end{document}